\documentclass[aps,pre,twocolumn,groupedaddress,floatfix,superscriptaddress]{revtex4}
\usepackage{dcolumn}
\usepackage{amsmath}
\usepackage{amssymb}
\usepackage{lipsum}
\usepackage{graphicx}
\usepackage{bm}
\usepackage[T1]{fontenc}
\usepackage{color}
\usepackage{url}
\usepackage[bf]{subfigure}
\usepackage{rotating}
\usepackage{soul}
\usepackage{scalefnt}
\usepackage{hyperref}
\usepackage{scalefnt}

\usepackage{perpage} 
\MakePerPage{footnote}

\usepackage{color}
\definecolor{redcolor}{rgb}{1.0,0.,0.}

\usepackage[utf8]{inputenc}
\begin{document}
\title{The Dynamics of Knowledge Acquisition via Self-Learning in Complex Networks}

\author{Thales S. Lima}
\affiliation{Institute of Mathematics and Computer Science, University of S\~ao Paulo, S\~ao Carlos, SP, Brazil.}
\author{Henrique F. de Arruda}
\affiliation{Institute of Mathematics and Computer Science, University of S\~ao Paulo, S\~ao Carlos, SP, Brazil.}
\author{Filipi N. Silva}
\affiliation{S\~ao Carlos Institute of Physics, University of S\~ao Paulo, S\~ao Carlos, SP, Brazil}
\affiliation{School of Informatics, Computing and Engineering, Indiana University, Bloomington, Indiana 47408, USA}
\author{Cesar H. Comin}
\affiliation{Department of Computer Science, Federal University of S\~ao Carlos - S\~ao Carlos, SP, Brazil}
\author{Diego R. Amancio}
\affiliation{Institute of Mathematics and Computer Science, University of S\~ao Paulo, S\~ao Carlos, SP, Brazil.}
\affiliation{School of Informatics, Computing and Engineering, Indiana University, Bloomington, Indiana 47408, USA}
\email{diego.raphael@gmail.com}
\author{Luciano da F. Costa}
\affiliation{S\~ao Carlos Institute of Physics, University of S\~ao Paulo, S\~ao Carlos, SP, Brazil}

\begin{abstract}
Studies regarding knowledge organization and acquisition are of great importance to understand areas related to science and technology.  A common way to model the relationship between different concepts is through complex networks. In such representations, network's nodes store knowledge and edges represent their relationships. Several studies that considered this type of structure and knowledge acquisition dynamics employed one or more agents to discover node concepts by  walking on the network. In this study, we investigate a different type of dynamics considering a single node as the ``network brain''. Such brain represents a range of real systems such as the information about the environment that is acquired by a person and is stored in the brain. To store the discovered information in a specific node, the agents walk on the network and return to the brain.  We propose three different dynamics and test them on several network models and on a real system, which is formed by journal articles and their respective citations. {Surprisingly, the results revealed that, according to the adopted walking models, the efficiency of self-knowledge acquisition has only a weak dependency on the topology, search strategy and localization of the network brain.
}
\end{abstract}

\maketitle

\noindent
``{\it The one self-knowledge worth having is to know one's own mind''}. F. H. Bradley


\setcounter{secnumdepth}{1}

\section{Introduction}
\begin{figure}[!htbp]
 \centering
 \subfigure[]{\includegraphics[width=0.90 \linewidth]{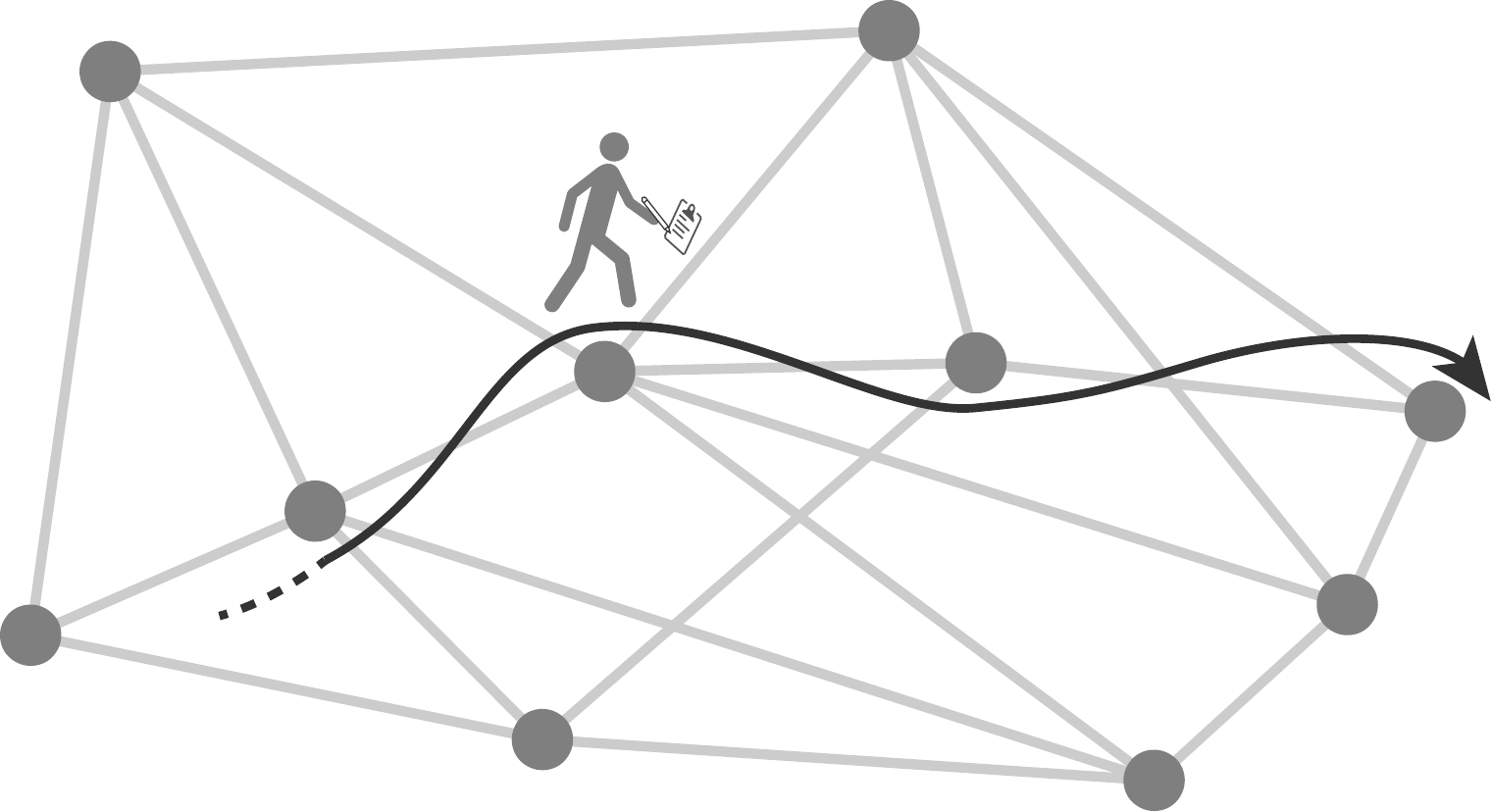}} 
 \subfigure[]{\includegraphics[width=0.90 \linewidth]{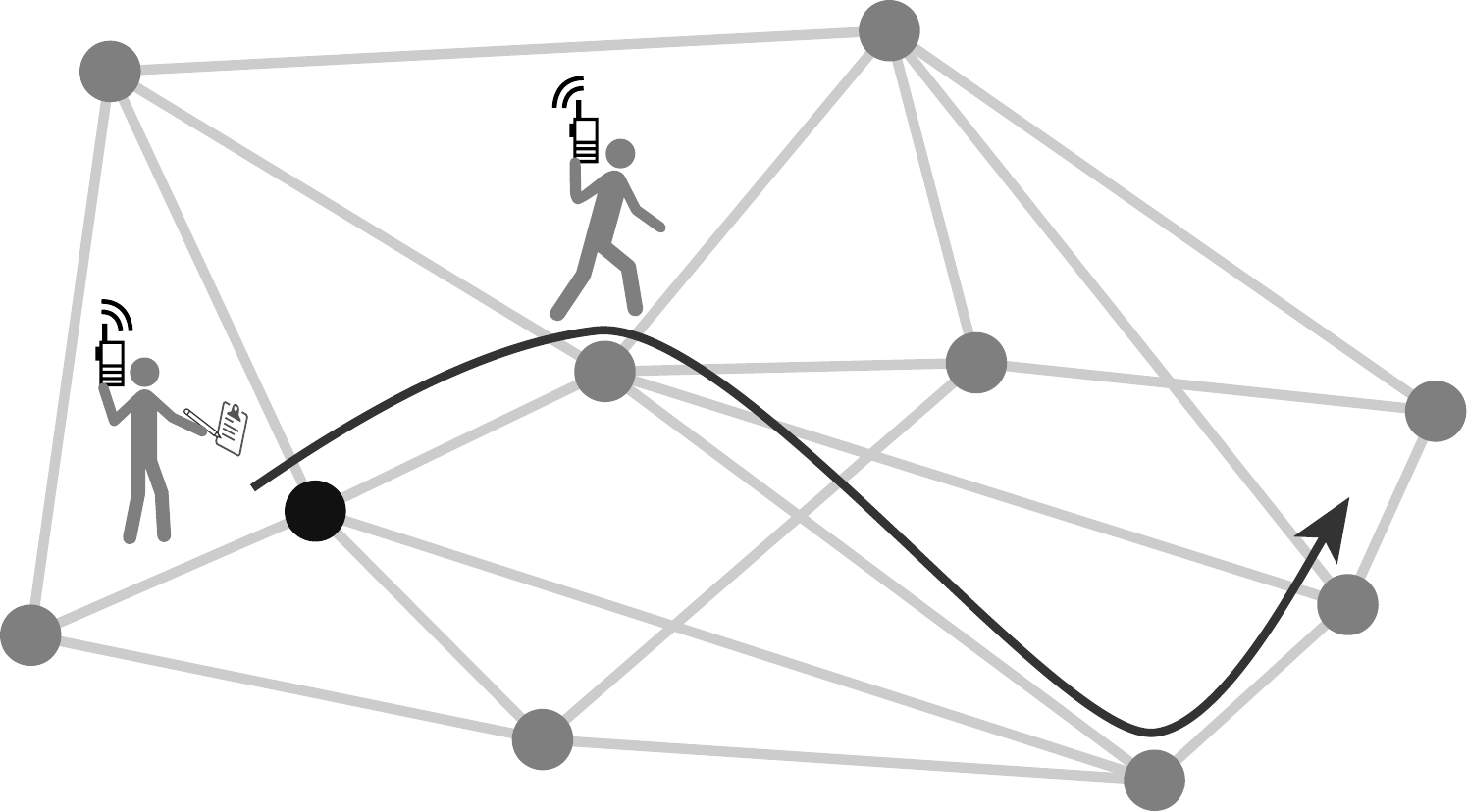}}
 \caption{Two possible cases of knowing a network.  In the first case (a), an agent repeatedly departs from different nodes and perform random walks aimed at collecting knowledge about the network topology.  In the second case (b), the collection of knowledge about the network is centered at a fixed node, here called \emph{network brain}.  An agent repeatedly departs from the \emph{same} node, collecting information about the network topology.  The latter situation is particularly interesting in the sense that the choice of the \emph{brain} can influence strongly the knowledge acquisition efficiency.}
 \label{fig:dynamics}
\end{figure}

Most matter and energy exchanges in nature can be represented and understood in terms of information~\cite{stonier1996information} flowing between distinct agents and/or subsystems.  Perhaps as a consequence, several developments in network science have shown that a great number of natural systems can be represented and modeled in terms of intricate graphs, which have been called \emph{complex networks}. Thus, interesting real-world systems such as opinion and epidemic spreading~\cite{pastor2001epidemic,moreno2004dynamics,pastor2015epidemic}, transport and communications~\cite{danila2006optimal}, as well as the nervous system~\cite{bullmore2009complex} have been approached by network science, yielding several interesting results.  One particularly interesting research line concerns how information is acquired by agents in a complex system~\cite{de2017connecting}.  Preliminary investigations along this line have represented knowledge as being stored into nodes that are logically interconnected, while one or more agents get acquainted with this knowledge while moving along the network~\cite{arruda2017knowledge}.  Therefore, the knowledge integration is performed internally to each agent.  This situation is illustrated in Figure~\ref{fig:dynamics}(a).

While many real-world situations --- such as web surfing and learning --- can be represented and modeled by using this agent-based approach, there are other systems underlain by a different dynamics.  The main difference in these cases regards the fact that the knowledge is integrated not by one or more agents, but is otherwise  stored in a specific and fixed part of the system, corresponding to some kind of `brain' or `hq', and the information is gathered by sampling through the networks by agents or sensing pathways.  Real-world related systems include the mapping of communities with shared interests in the WWW or the Internet, as well as consciousness acquisition by individuals with respect to environment. In these situations, the information to be integrated can be considered as being part of the own organism, so that acquired knowledge becomes \emph{self-knowledge}.  For simplicity's sake, the central node where information is integrated in these systems is henceforth called the \emph{network brain}.

This type of centralized knowledge acquisition, illustrated in Figure~\ref{fig:dynamics}(b), has seldom been addressed from the perspective of computational modeling.  Yet, this type of systems entails some specific problems that are particularly interesting.  One of the main such problems regards how influential is the topology around the brain on the knowledge acquisition speed.  In case the topology is influent, this would imply in special care while electing a specific node to operate as brain or hq of the system whenever possible.  The present work addresses precisely this type of centralized systems and influence of topology in brain placement.  We resort to complex network models and concepts in order to investigate how self-knowledge progress in different types of networks and how this dynamics is influenced by the choice of node to act as the network brain. 

In the investigations reported in this work, we consider the domain to be explored to be represented as a complex network. The knowledge exploration is performed by agents that move along the network through random walks and periodically report the information gathered to the network brain, where it is integrated and stored. {These agents can be thought as independent and simultaneous, thus the knowledge of different walks is not transmittable between them}.  A number of different topologies are considered in order to allow us to infer to which extent the structure of the domain influences the self-exploration efficiency.  At least as important as the topology in defining such an efficiency, is the choice of node to act as brain and the type of probing dynamics.

Recent studies suggest that the topology of the network, and also around the starting node, tends to have little influence on the knowledge acquisition speed, at least for the considered models~\cite{arruda2017knowledge}.  However, these results refer to moving agents as in Figure~\ref{fig:dynamics}.  Because the random walks implemented in such cases tend to `forget' the initial node, the topology around these nodes are intrinsically less influent on the knowledge acquisition.   In the current work, however, the node where the knowledge is integrated is fixed, and all the several implemented random walks depart from this same node.  Therefore, it could be expected that the topology surrounding these nodes play a more decisive role on the knowledge acquisition speed.  


The remainder of this article is organized as follows. Section~II presents the employed network and dynamics. In section~III, we analyze the adopted dynamics and compare their efficiency according to different used parameters. Finally, in Section~IV we conclude the study and describe some remarks regarding the results. 

\section{Materials and Methods}

This section presents the adopted networks, which include a real-world case and several theoretical models, the topological measurements used to characterize their topology, and the implemented self-learning dynamics.

\subsection{Complex Networks Models}

Networks with the same size and number of edges were then for each different model studied. The models that were studied in this work were:

\emph{Random Networks (Erdős-Rényi -- ER):} This has been used as the reference model in complex networks research.  Each of its nodes has a probability $p$ of connecting to each other node in the network~\cite{erdos1960evolution}.

\emph{Scale Free Networks (Barabasi-Albert -- BA):} This type of network is generated by incrementally adding nodes with probability of connection  proportional to the degree of the nodes already existing in the network. The most import property of this network model is the degree power law and the hence implied presence of hubs~\cite{barabasi1999emergence}.

\emph{Configurational Model (CM):} In this model, a random graph is created to fit a desired degree distribution~\cite{newman2003structure}. The confrontational model was used as a means to better understand some complex effects detected in the BA networks. The degree distributions used in this work were generated to follow a BA network.

\emph{Small World Networks (Watts-Strogatz -- WS):} This model is generated by rewiring connections in a regular toroidal lattice. It presents the combination of high clustering coefficient and low average minimal distance~\cite{watts1998collective}. The standard rewiring probability used was $3\%$.

\emph{Geographical Networks (Waxmann -- Wax):} These are based on distributing points in the Cartesian plane. The vertices are evenly distributed, and each node can connect to each other with probability decreasing exponentially with the distance.  For the Waxmann  networks we used a square space with random distribution o vertices in it~\cite{waxman1988routing}.

\emph{Stochastic Block Model (SBM):} This model uses blocks of nodes and probabilities of connection between nodes from one block to another (even with itself)~\cite{batagelj2004generalized, holland1983stochastic}. Two models of SBM networks were adopted in this work, both with 10 blocks. Each block has the same probability $\mu$ of connecting to the other blocks, the first model has a $\mu = 1\%$ chance and the second $\mu = 2\%$ chance.

All parameters referring to connection between nodes were chosen to approximately reflect the desired average degree of the network, i.e. the probability of connection between two node in the ER network. 

\subsection{Real-World Network}
In order to investigate the behavior of the proposed dynamics on a real system that represent knowledge relationships, we adopted a citation network obtained from the Web of Science (WOS)~\footnote{\url{http://www.webofscience.com}}. The network comprises articles returned by searching for complex network topics in journals indexed by WOS. Each node corresponds to an article and a connection exists between a pair if they cite each other. Note that, in this case, we disregard the direction of the connections since we are interested in the relatedness between topics and not in citation dynamics. The network contains $11,063$ articles and has average degree $17$. More information about this network can be found in \cite{silva2016using}.

\subsection{Walk Models}

With these networks in hand, we aim at understanding how long does it take for a random walker to fully discover the  network. Three different walking models were used:

\begin{figure*}[!b]
  \centering
    \includegraphics[width=.8\textwidth]{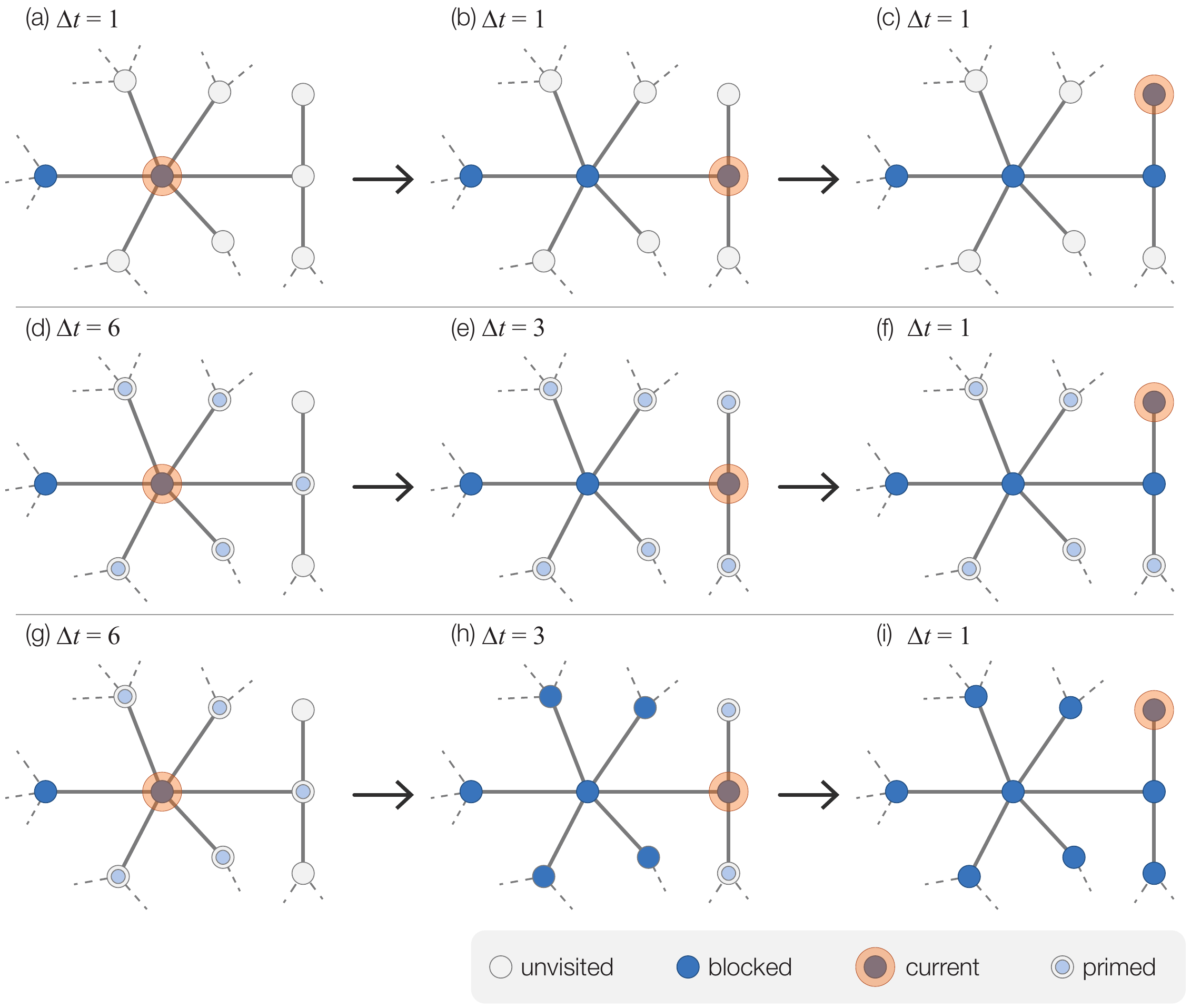}
   \caption{Description of the changes of states in the three adopted walking models.  The four states that a node can assume are shown in the legend, and include: \emph{unvisited}, \emph{primed} (a node that is known but can still be visited), \emph{blocked} (a node that is known and cannot be visited), and \emph{current} (the node where the agent is). (a-c) Represents the self avoiding dynamics, where the only known nodes are those that have been visited already, (d-f) the extended self avoiding where all nodes that neighbor the path are also known but not avoided and (g-i) the look-ahead self avoiding where all nodes that neighbor the path are known and are avoided.}
  \label{WalkTypes}
\end{figure*}

\emph{Standard Self-avoiding random walks}: for every move in these walks, the walker chooses the destination node randomly in the current neighborhood. The sole restriction on these walks is that the same node cannot be visited twice. A node is considered to be known if the walker has visited it at least once. Therefore the number of steps of a walk using this dynamic is the number of nodes of the walk. Thus the number of steps of each iteration, denoted by $\Delta t$ is always $1$, as illustrated in Figure~\ref{WalkTypes}(a-c).

\emph{Extended self-avoiding random walks}: in this model the walker behaves exactly like the previous model, but the known region is considered to include not only the nodes that were covered along the walk, but also the neighbors of these nodes. This means that the number of steps in a walk correspond to the sum of the degrees of each of its nodes, as seen in Figure~\ref{WalkTypes}(d-f).

\emph{Look-ahead self-avoiding walks}: like the previous model, in this case the walker will be considered to know the neighborhood of every visited node after departing from it. However, the agents are prohibited to proceed to any already known node. The number of steps of a walk is calculated in the same fashion as in the extended walk dynamics. This dynamics is illustrated in Figure~\ref{WalkTypes}(g-i).

Note that, for every model, the walks may eventually hit a dead end and will have to stop, prompting another walk to be started from the the same starting point. This is repeated until all nodes have been discovered. For every new iteration the same starting node is used and the network is considered to be completely unvisited by the agent (the knowledge about the previous walks is always kept in the network brain).

Since every network model was created with the same number of vertices, but the average degree varies from 3 to 30, the probability of dead ends will vary considerably. To better grasp the influence of very long walks, an additional configuration was adopted in which the maximum number of steps before reseting the walk was set to 100.

\section{Results and Discussion}

The results in this section all refer to the number of steps that had to be taken in order to discover a given percentage of the network. The total number of steps is considered to be the sum of steps of all the walks. This was done to emulate the cost of acquiring knowledge about the network. 

\begin{figure*}[!p]
\begin{tabular}{ccc}
  \includegraphics[width=.33\textwidth]{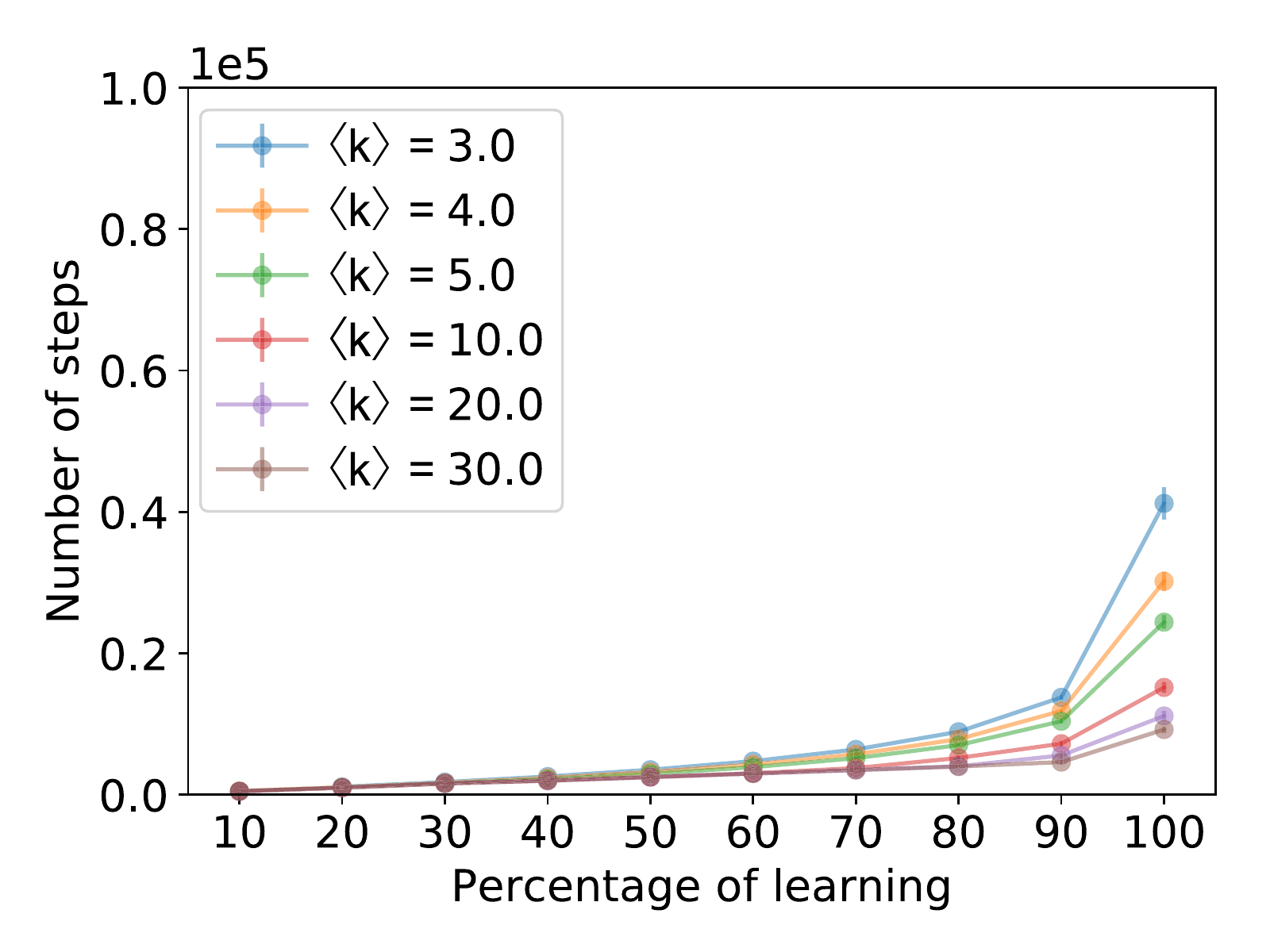} &   
  \includegraphics[width=.33\textwidth]{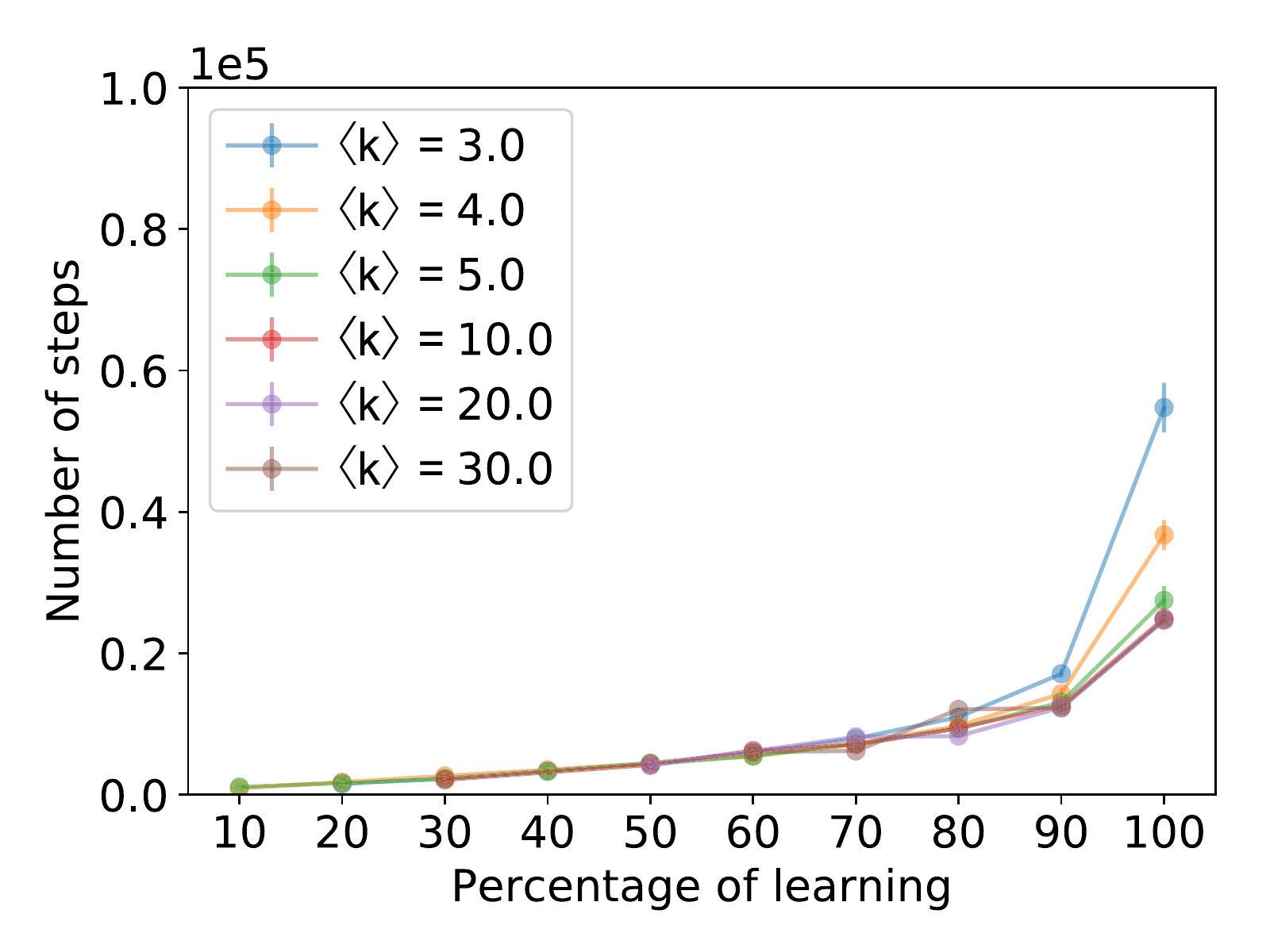} & 
  \includegraphics[width=.33\textwidth]{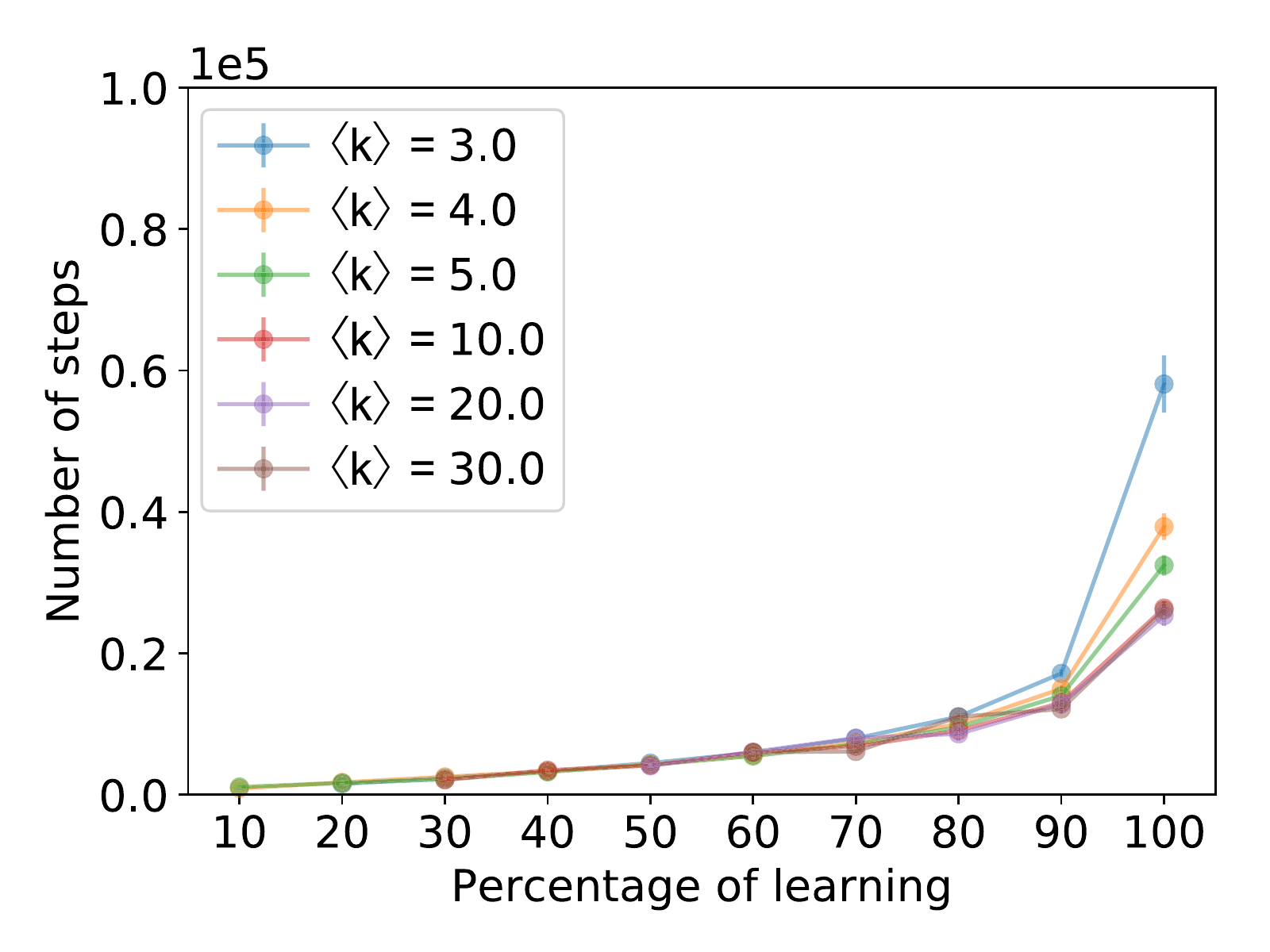}\\
  (a) Standard  & (b) Extended & (c)  Look-ahead\\[6pt]

\end{tabular}
\caption{Learning efficiency for the ER networks considering the three adopted dynamics. Each curve corresponds to a realization for a different average node degree $\langle k \rangle$, according to the legend.}
\label{erdos}
\end{figure*}

\begin{figure*}[!htb]
  \begin{tabular}{ccc}
  \includegraphics[width=.33\textwidth]{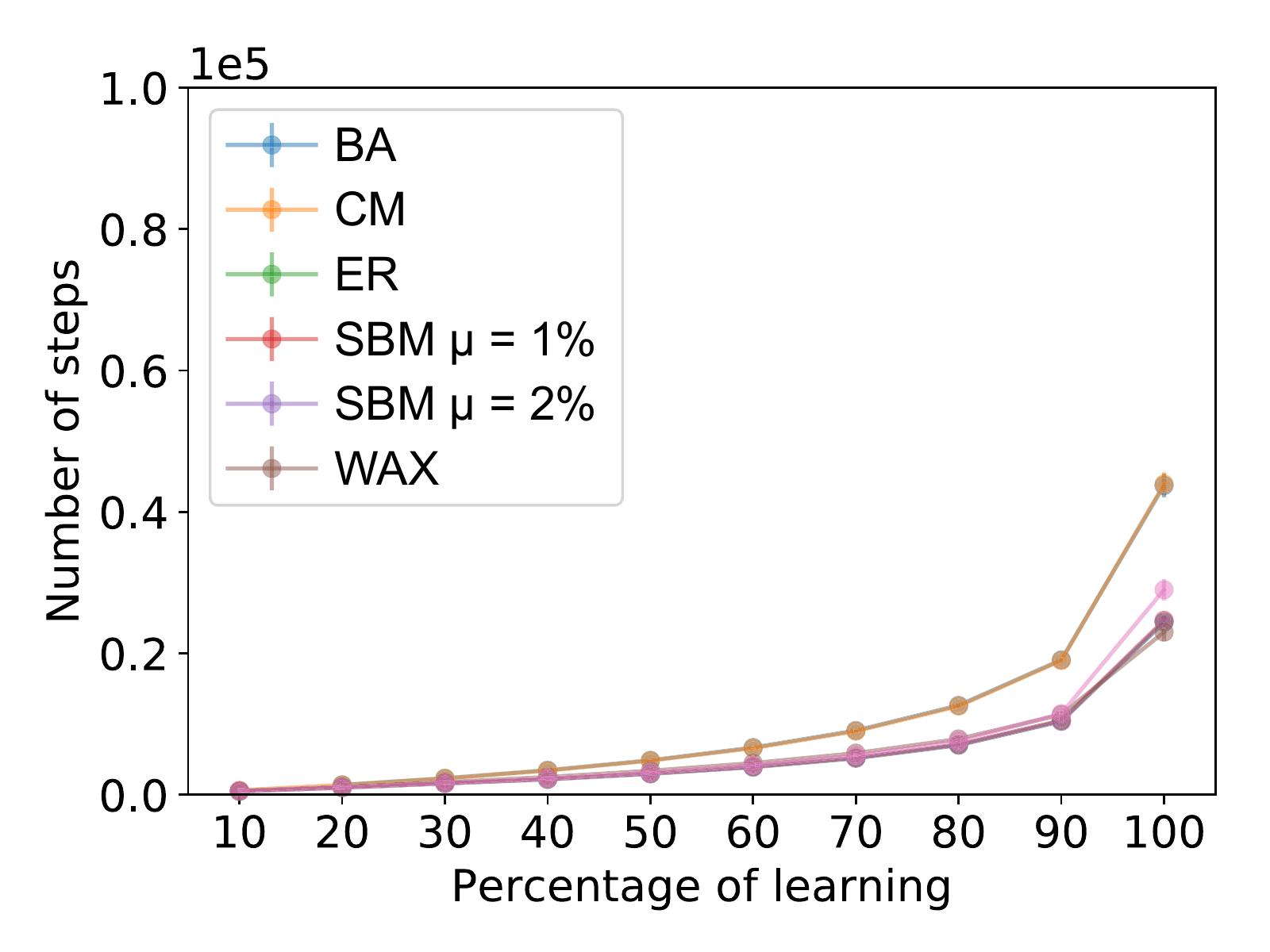} &   
  \includegraphics[width=.33\textwidth]{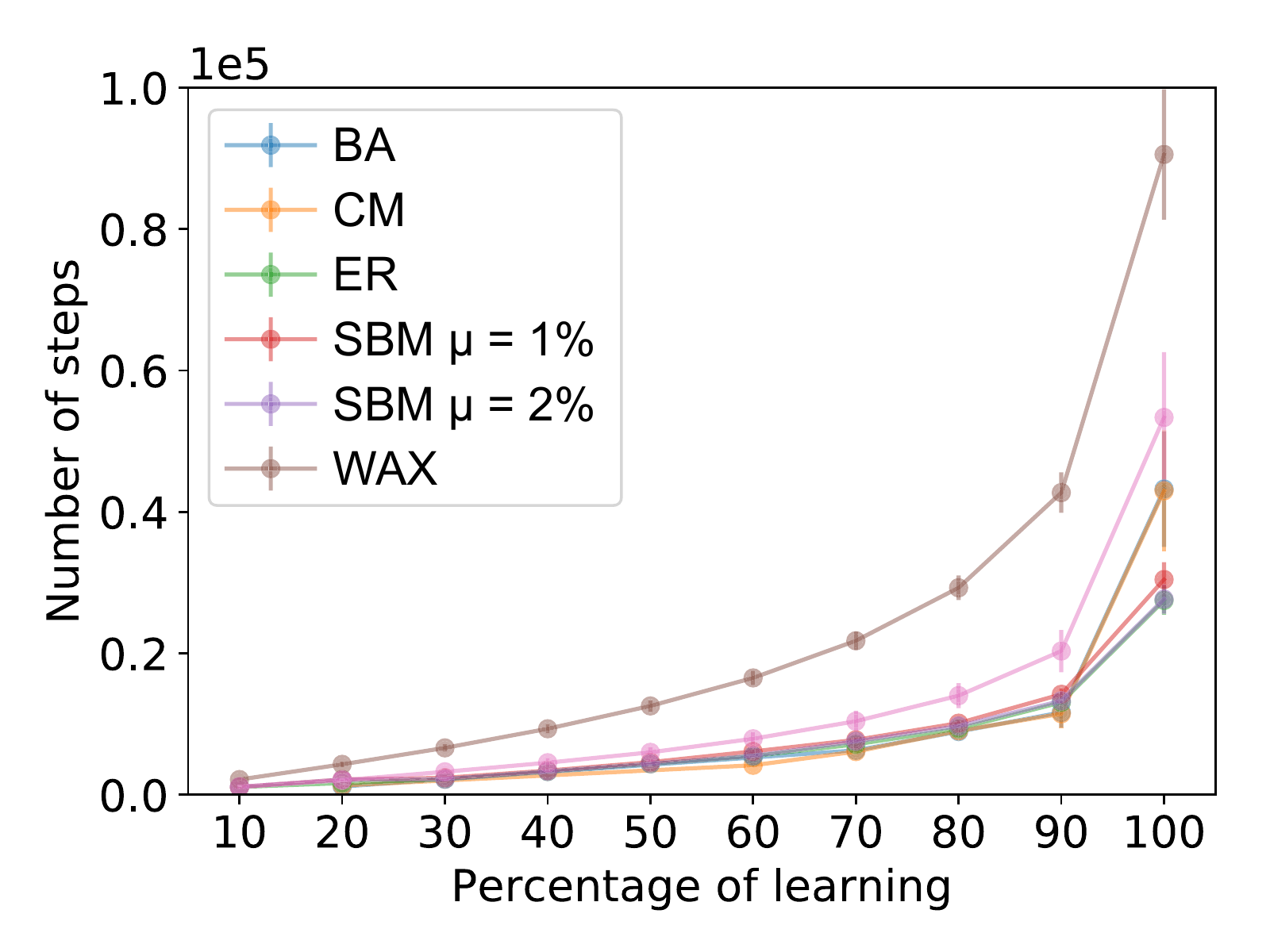} & 
  \includegraphics[width=.33\textwidth]{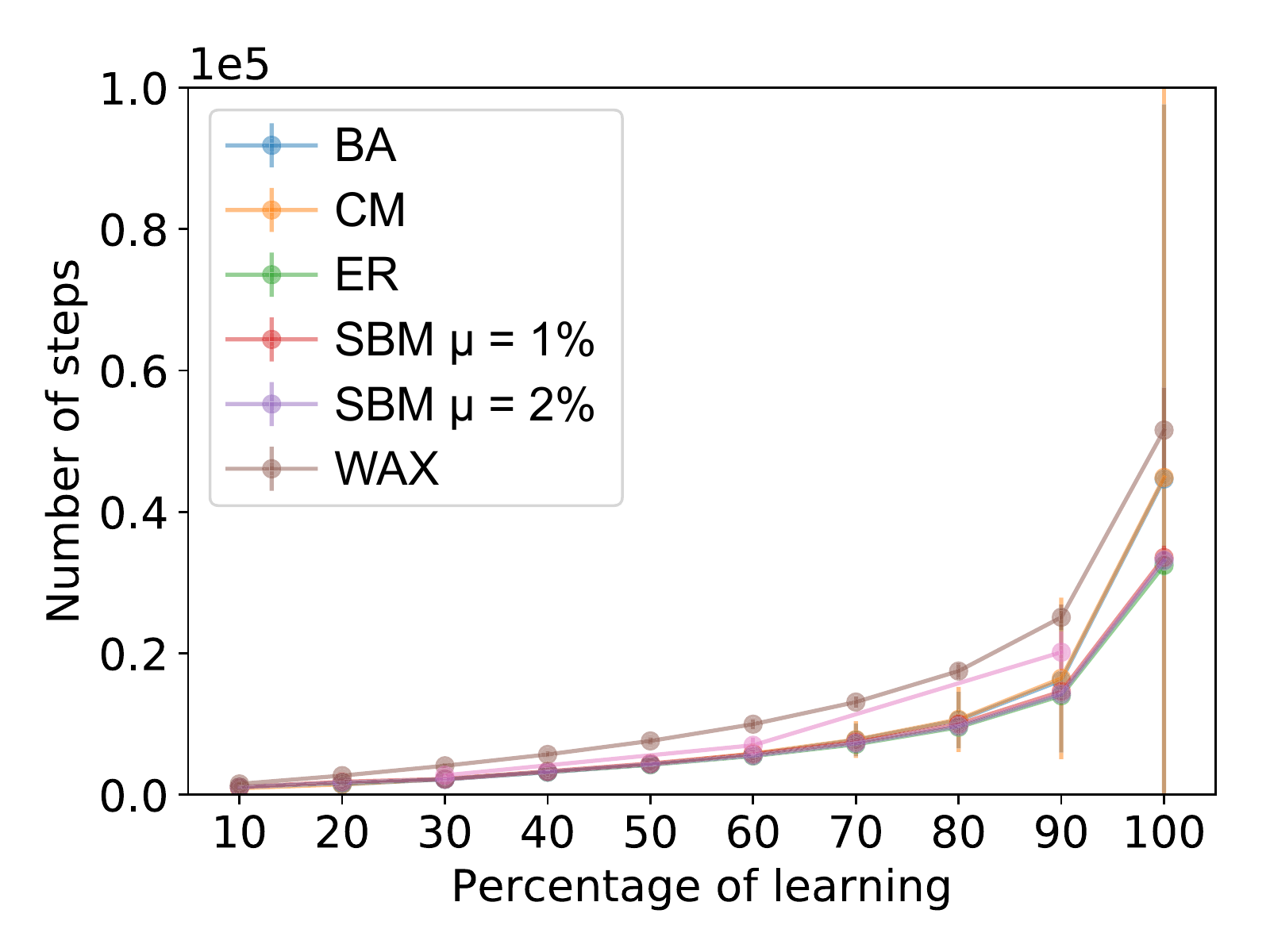}\\
  (a) Standard  & (b) Extended & (c)  Look-ahead\\[6pt]
\end{tabular}
   \caption{Efficiency curves for the considered walk dynamics and network models. In all cases, $\langle k\rangle = 5$.}
  \label{k_3}
\end{figure*}

The experiments were performed considering 100 different respective starting points in each network. First, the nodes are ordered according to their degrees. Next, the selected starting points are taken every 50 nodes in this sequence, implying in $2\%$ sampling of the nodes. This was done to better understand the influence of the centrality in the network on the total number of step necessary to discover the whole network.

\subsection{ER Network discovery}

The studied ER networks are composed by 5000 nodes with average degree $\langle k \rangle$ varying from $3$ to $30$. In Figure~\ref{erdos}, we show the number of steps required for covering a given percentage of the network, with respect to the three considered dynamics. A prototypic pattern can be identified in all cases, the only difference being an overall decrease with the average degree of the network.

Interestingly, there are no major differences between the extended self-avoiding and look ahead self avoiding types of walks. Avoiding or not the neighborhood of the walk bears little influence on the efficiency. 

\subsection{Other models}

When compared to other models we notice that the pattens displayed on ER networks are also present on almost all other models, this can be seen Figure \ref{k_3}. 

Some interesting behaviors have been identified. First, almost every considered network model and walk model combinations present the same decreasing pattern observed for the ER networks. The most deviating curves refer to the BA and CM models, which present a significant standard deviation on look-ahead walks. WS networks also present a slight variance in efficiency, especially for the extended walk. Such small changes are probably caused by the existence of a few shortest paths compared to the regular lattice (for this model, we chose to have only 3\% of the edges being rewired). This effect is more pronounced in the extended dynamic because it combines the two deficiencies from the other methods: it takes paths that are analogous to the normal dynamic, making it harder to reach distant part of the network, and it always checks the neighborhood of the walk, causing the nodes in the vicinity of the brain to be checked several times.

\subsection{Influence of the Starting Node}

\begin{figure*}[!htpb]
    \begin{tabular}{ccc}
  \includegraphics[width=.33\textwidth]{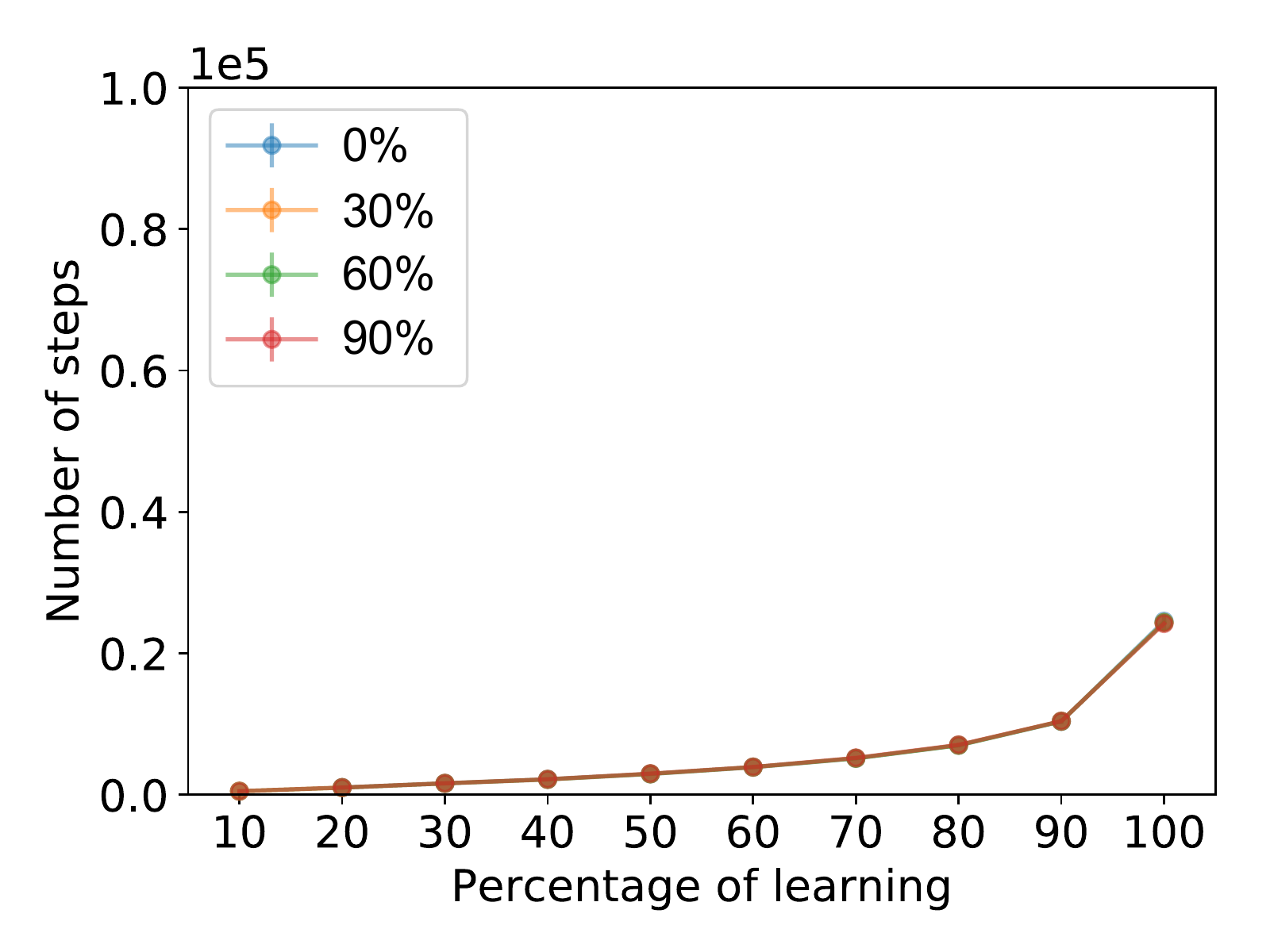} &   
  \includegraphics[width=.33\textwidth]{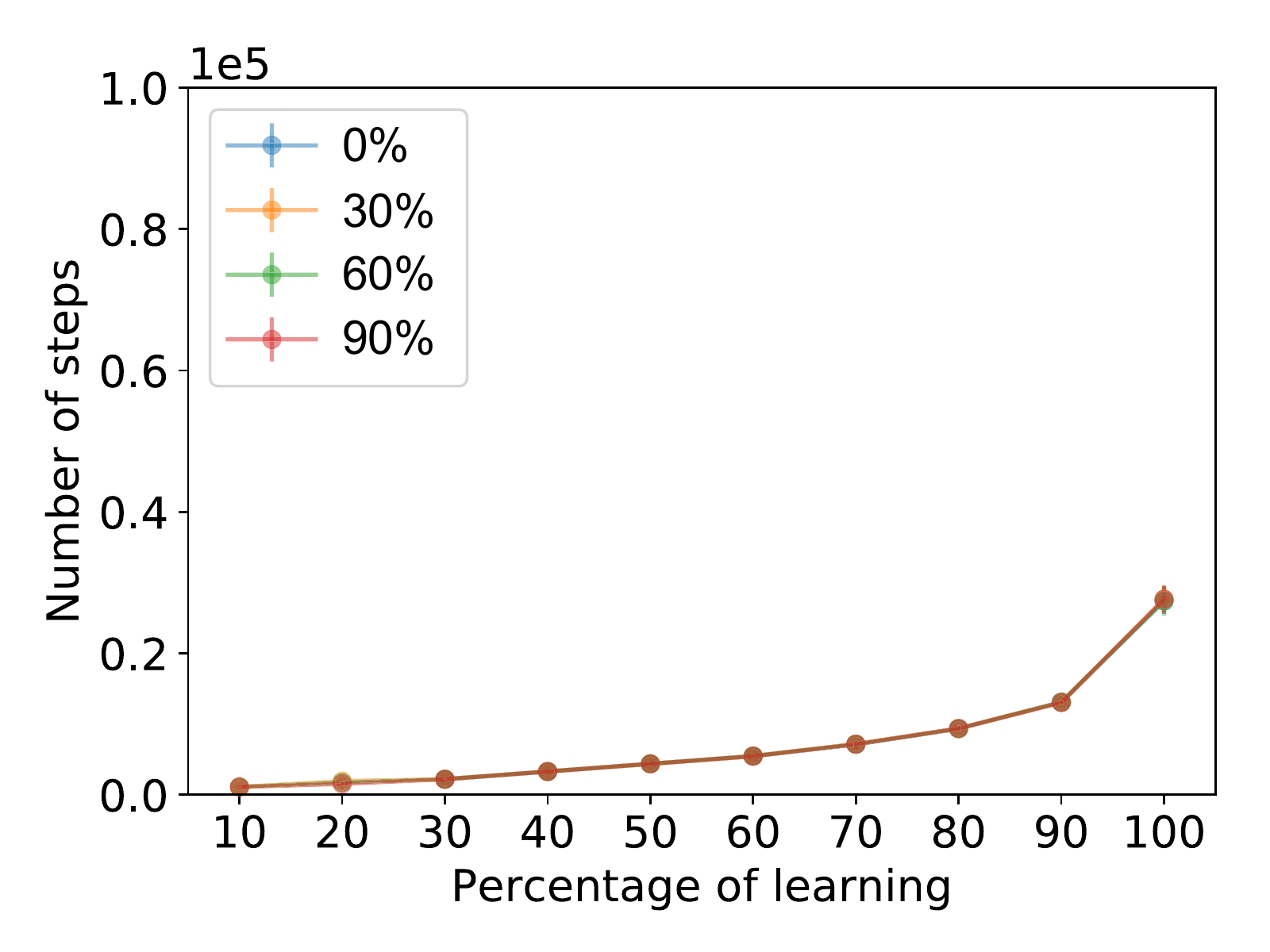} & 
  \includegraphics[width=.33\textwidth]{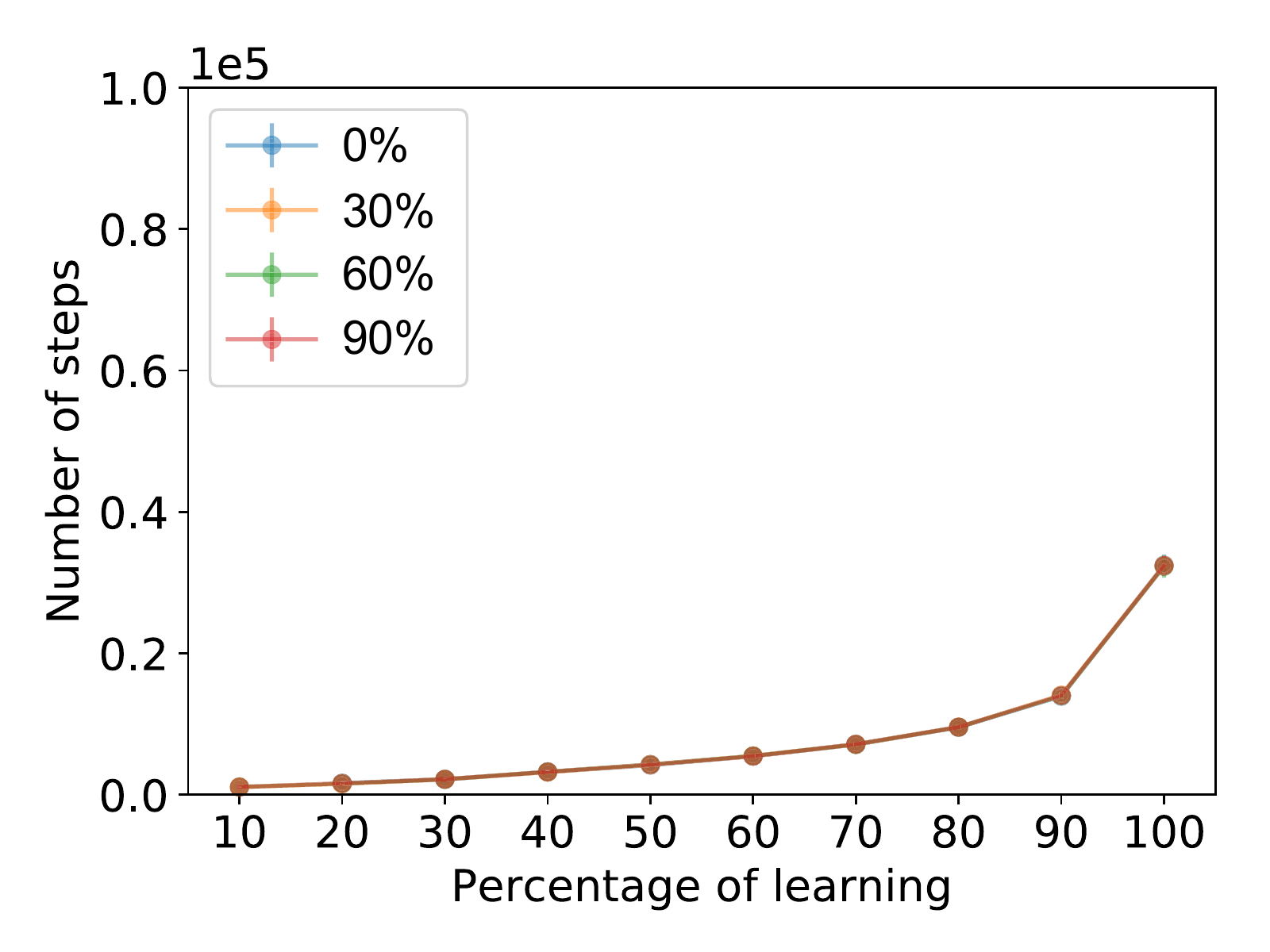}\\
  (a) ER Standard  & (b) ER Extended & (c)  ER Look-ahead\\
   \includegraphics[width=.33\textwidth]{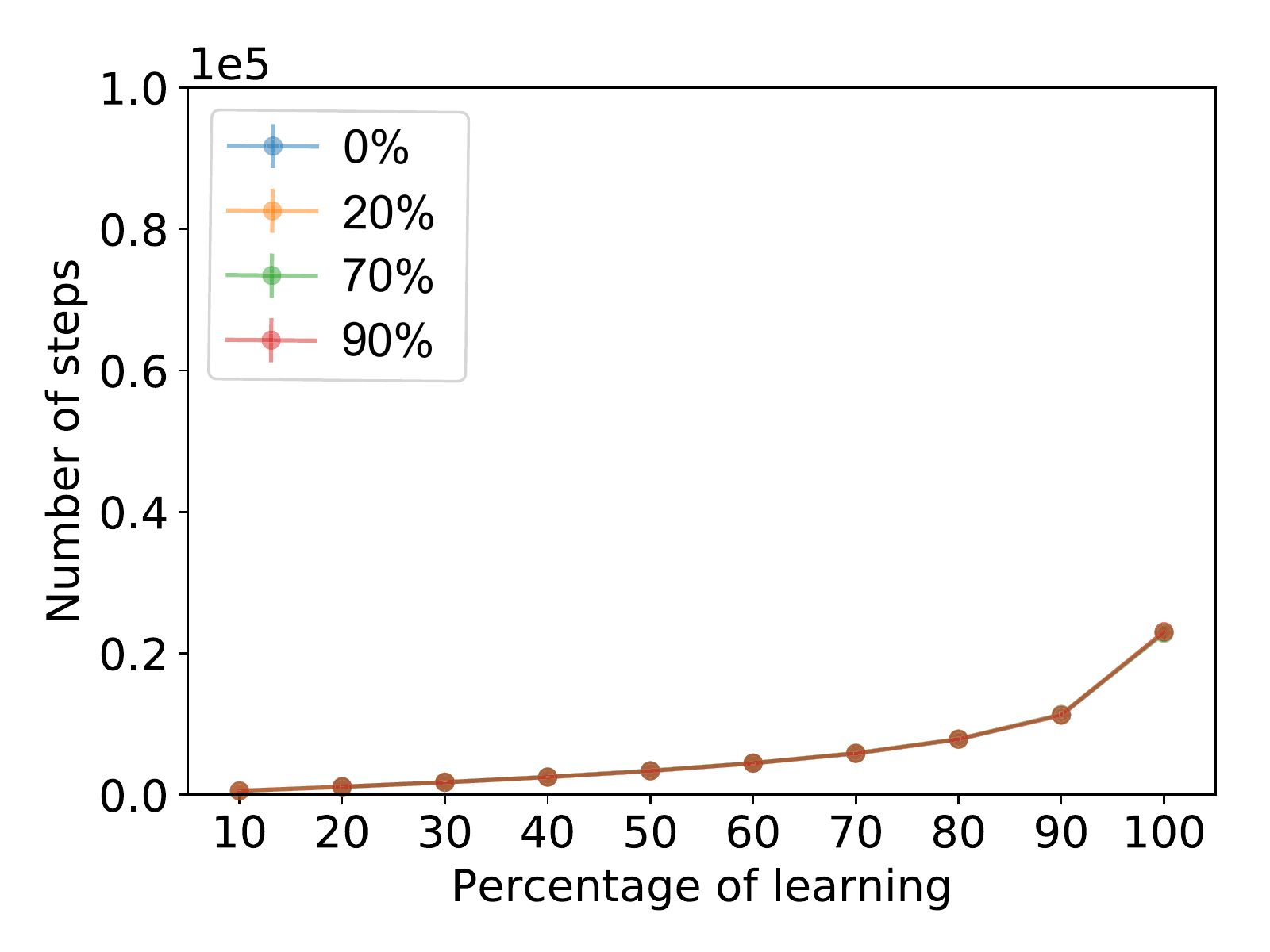} &   
  \includegraphics[width=.33\textwidth]{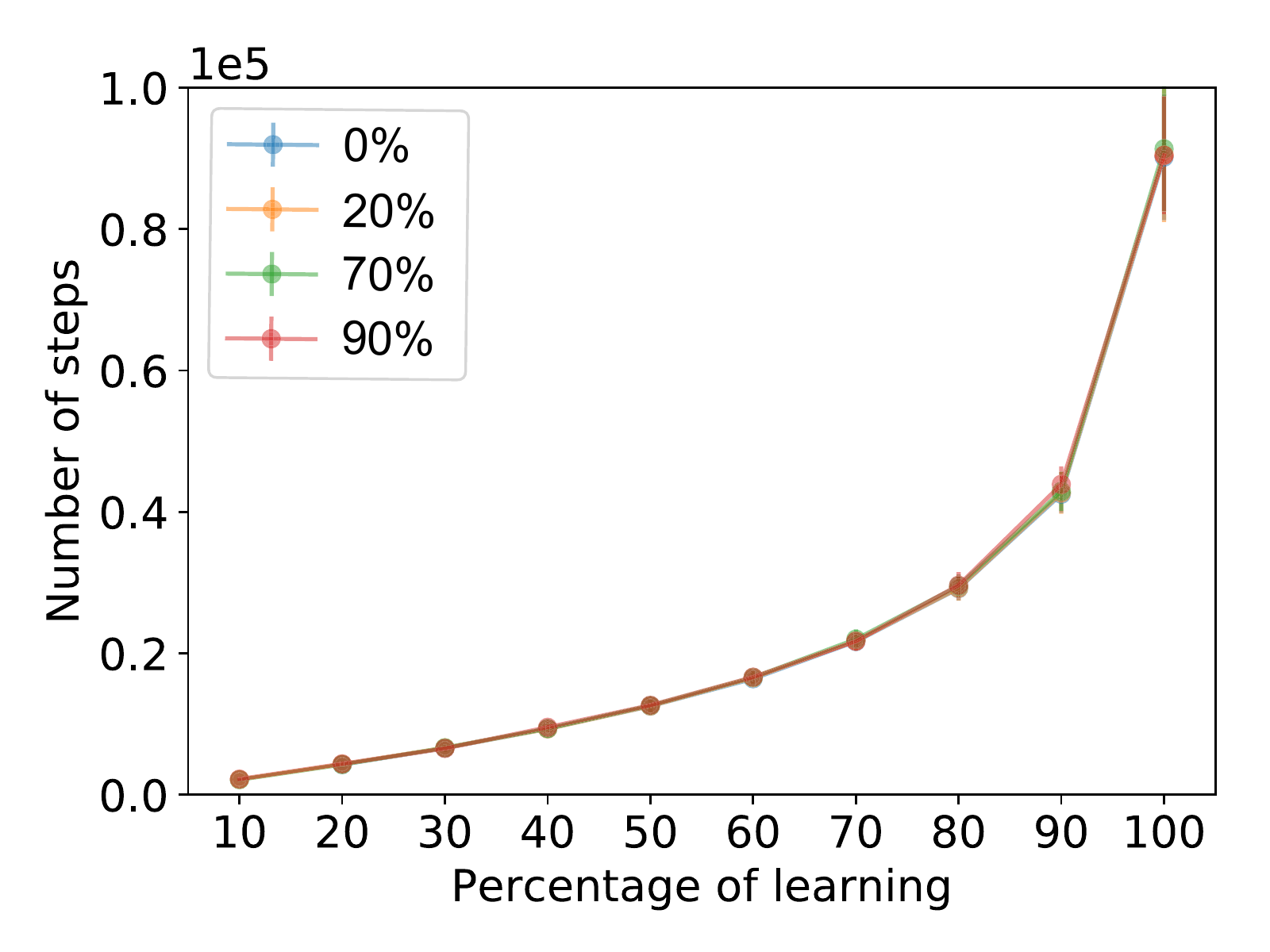} & 
  \includegraphics[width=.33\textwidth]{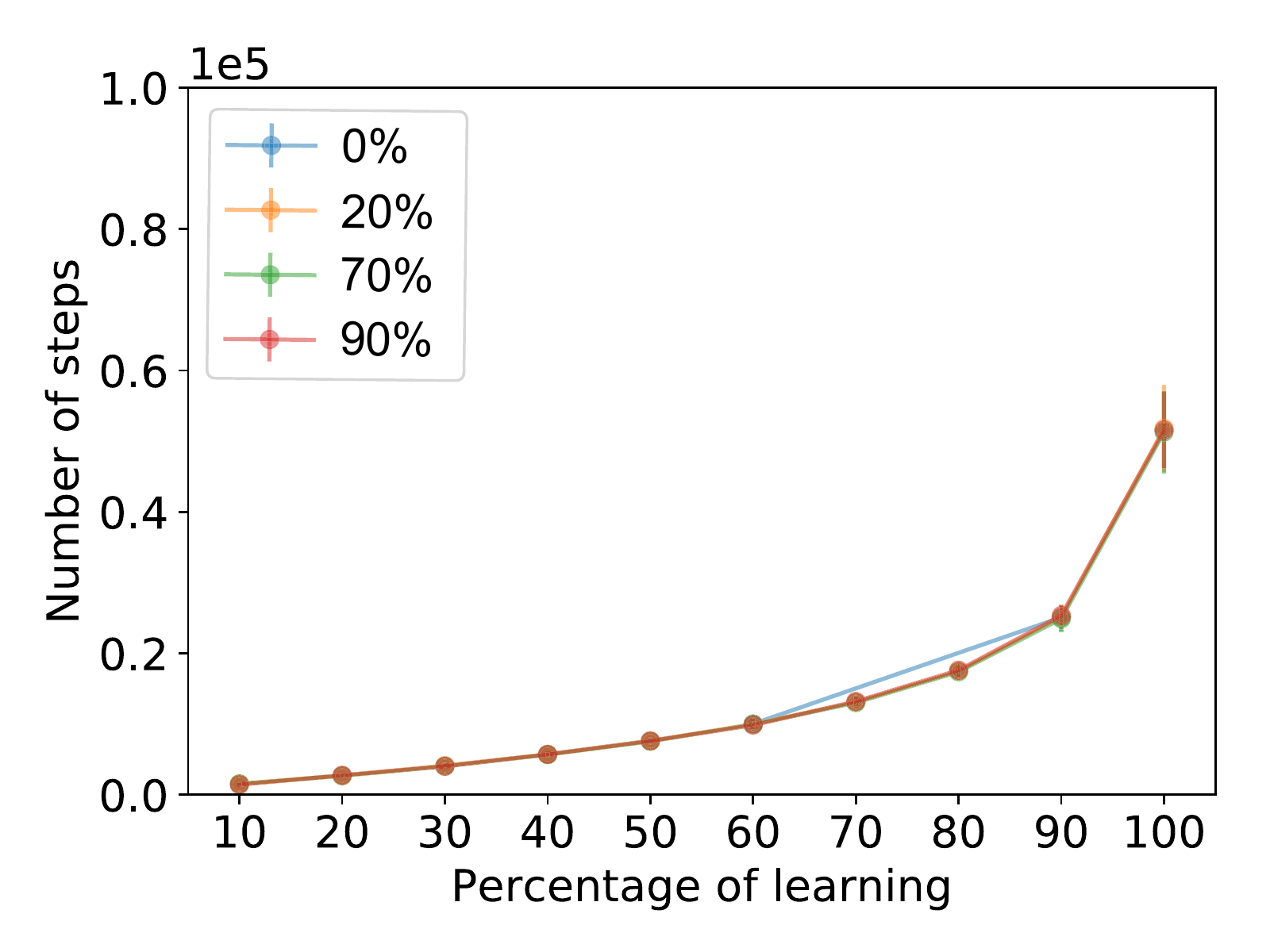}\\
  (d) WS Standard  & (e) WS Extended & (f)  WS Look-ahead\\
  \includegraphics[width=.33\textwidth]{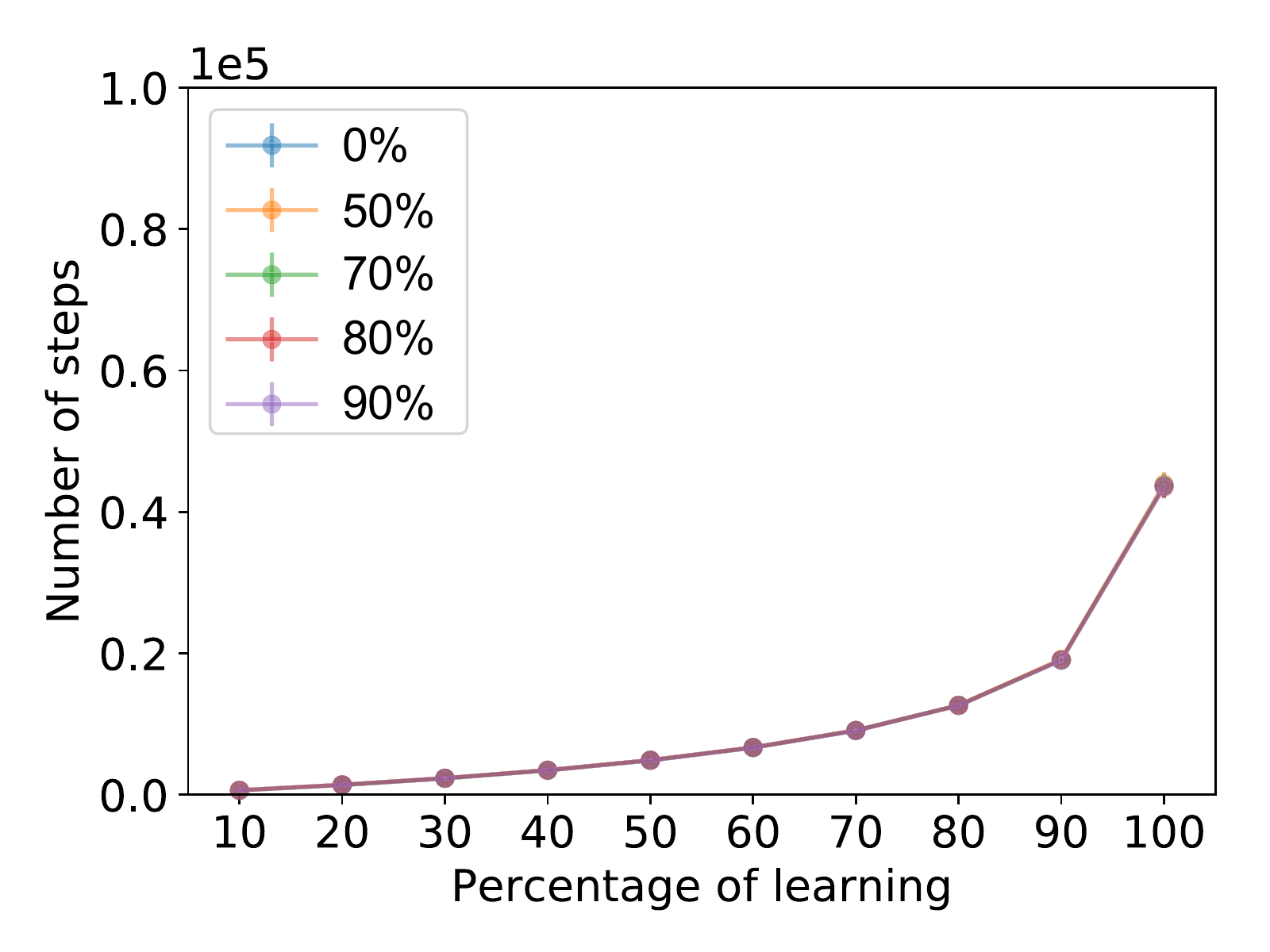} &   
  \includegraphics[width=.33\textwidth]{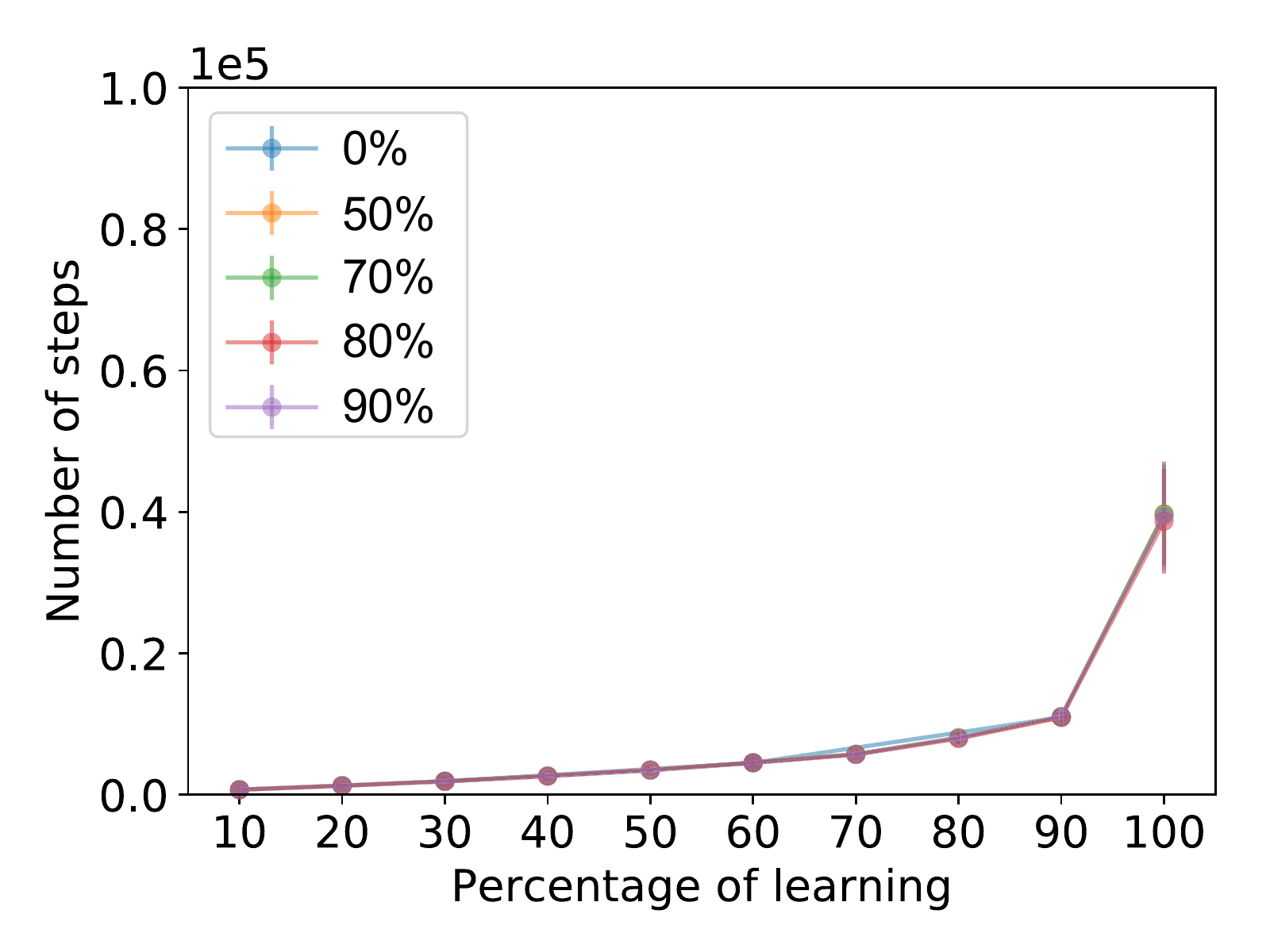} & 
  \includegraphics[width=.33\textwidth]{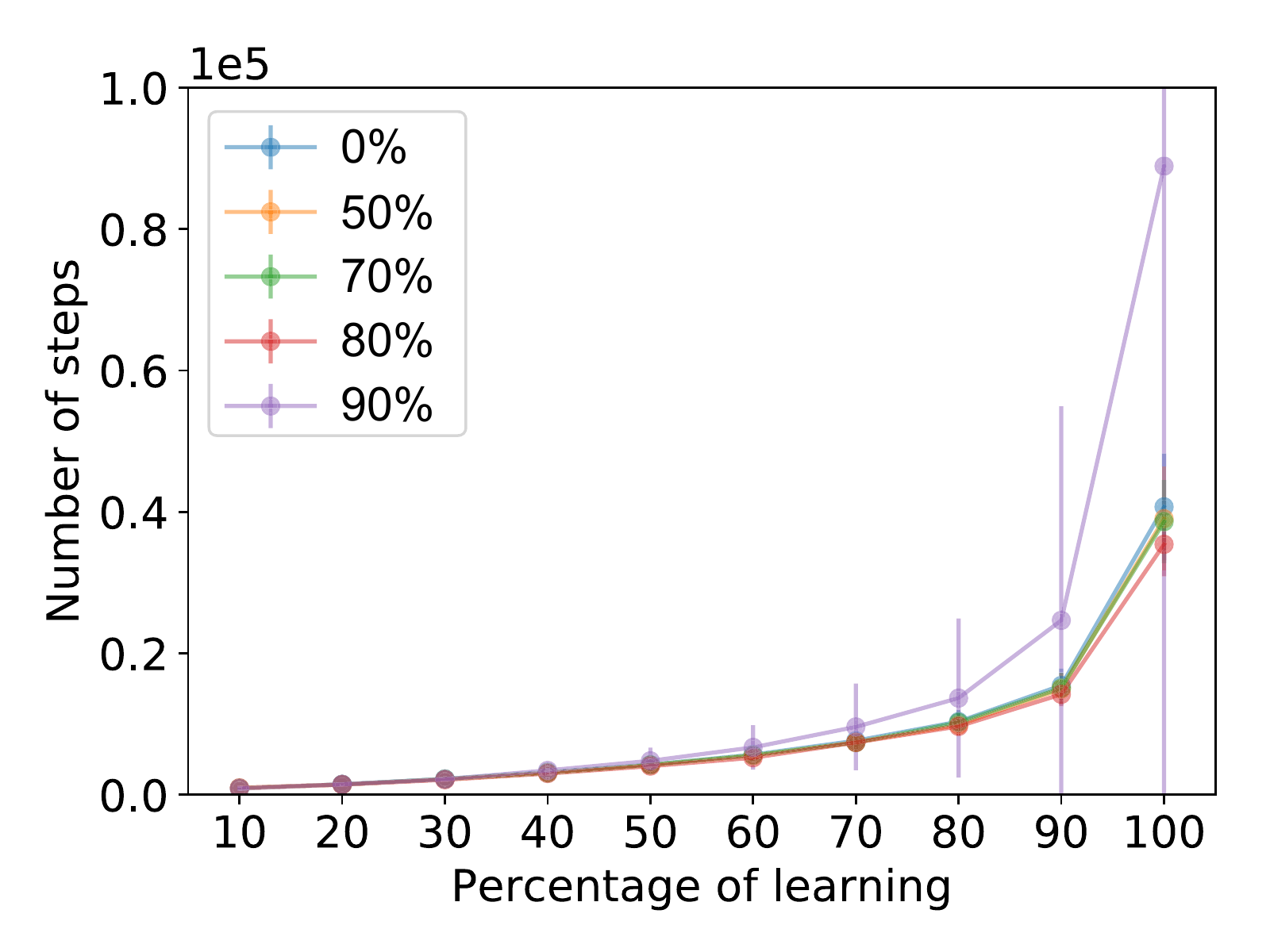}\\
  (g) BA Standard  & (h) BA Extended & (i)  BA Look-ahead
\end{tabular}
   \caption{Learning efficiency obtained for all the considered combinations of walk dynamics and network models. Each curve corresponds to a selection of starting node having betweenness centrality higher than a certain percentile considering the entire network, as indicated in the legends.}
  \label{centrality}
\end{figure*}

We also studied the influence of the centrality of the chosen starting node on the total number of steps required for exploring the whole network. We found that the centrality (computed via degree and betweenness) does not have great influence on the final results, with the only exception being the BA networks when considering the look-ahead self avoiding walks (Figure \ref{centrality}). Notably, this influence is only significant when the node of origin is a hub.  In order to understand if this is a property specific to BA networks, we compared this model with CM networks with the same degree distribution.  The same behavior was observed, leading us to believe that the actual degree distribution has a more important role in this dynamics than the model itself. 

\begin{figure*}[!htb]
\vspace{1cm}
  \begin{tabular}{ccc}
  \includegraphics[width=.33\textwidth]{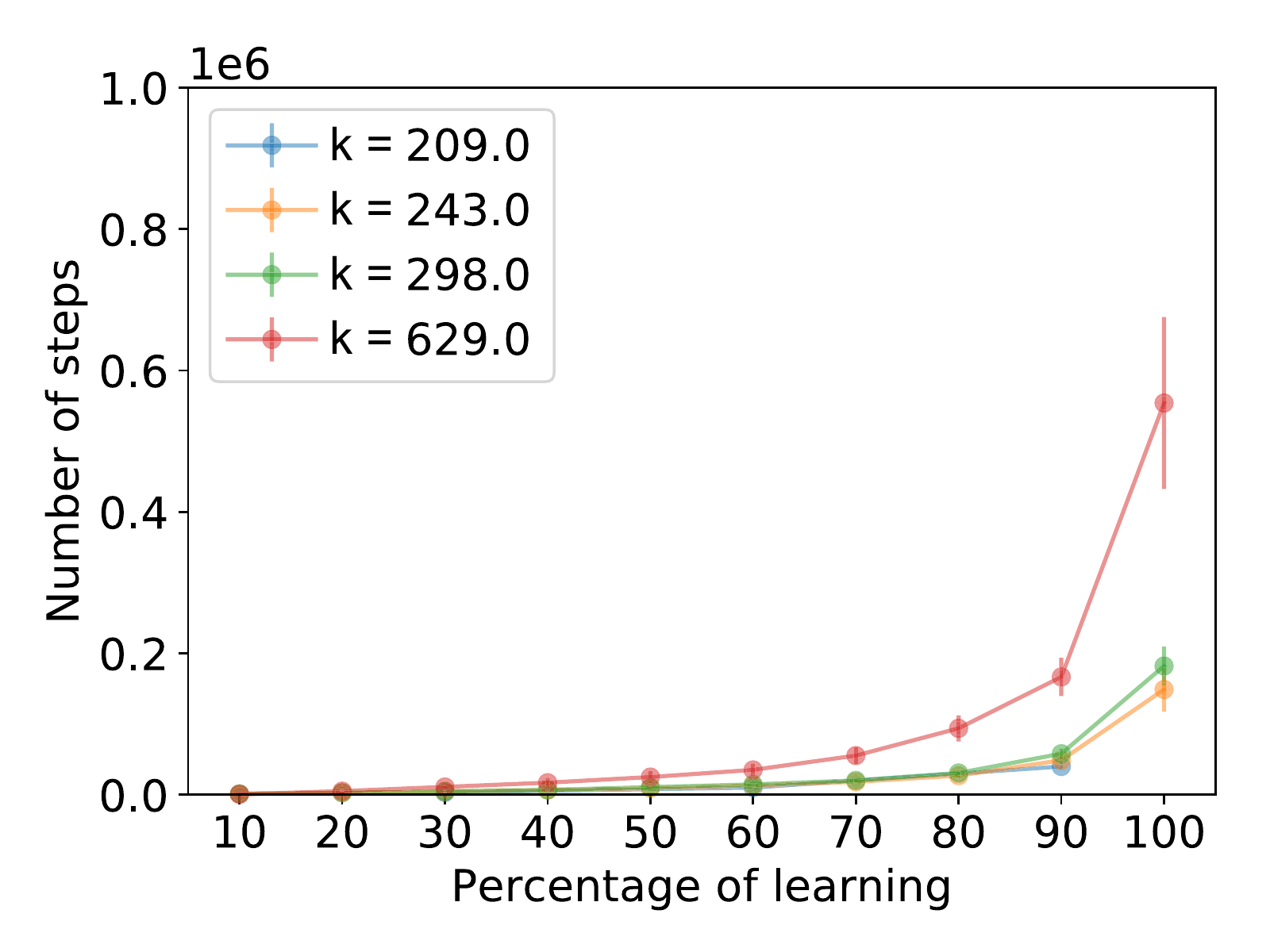} &   
  \includegraphics[width=.33\textwidth]{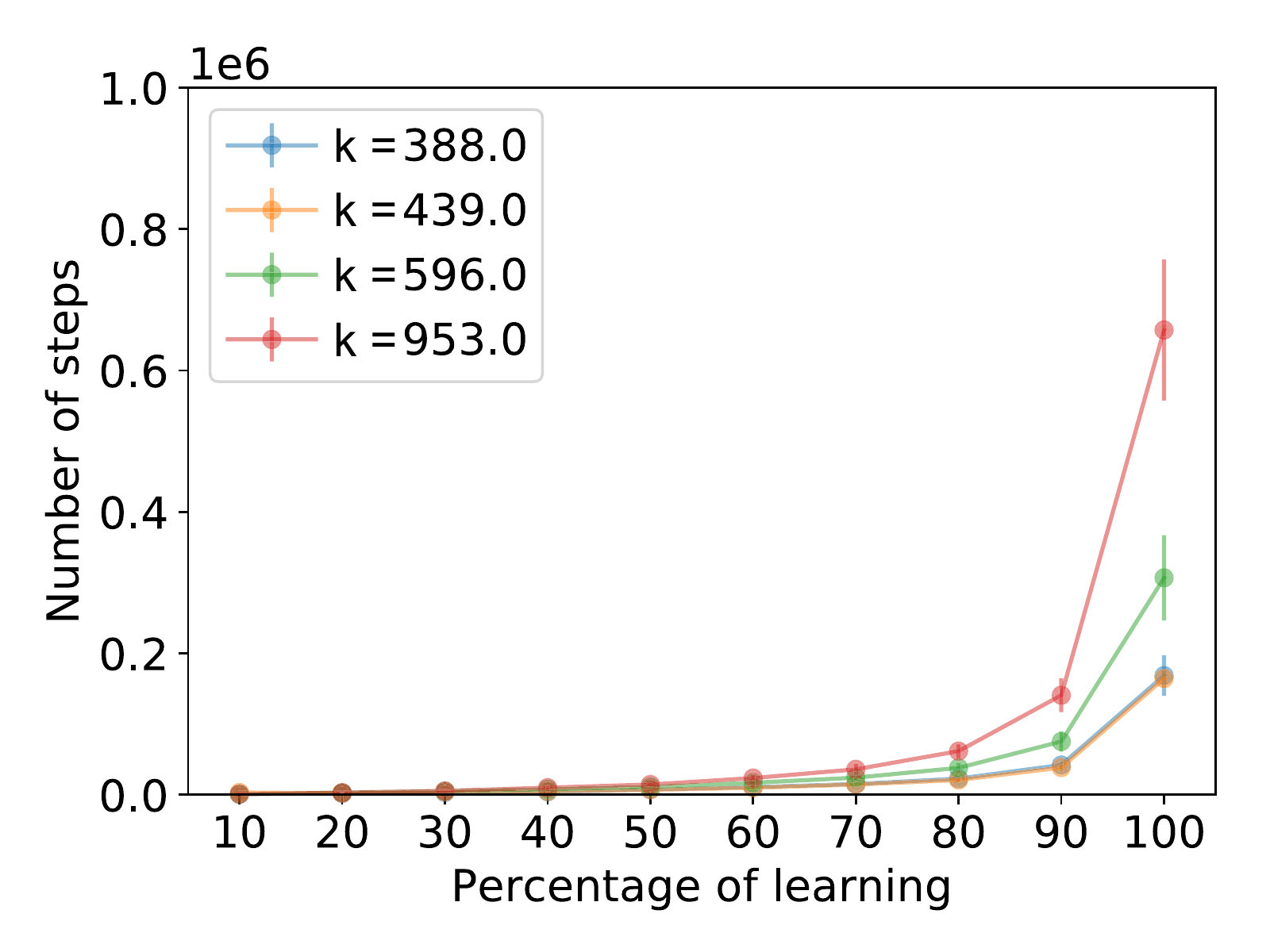} & 
  \includegraphics[width=.33\textwidth]{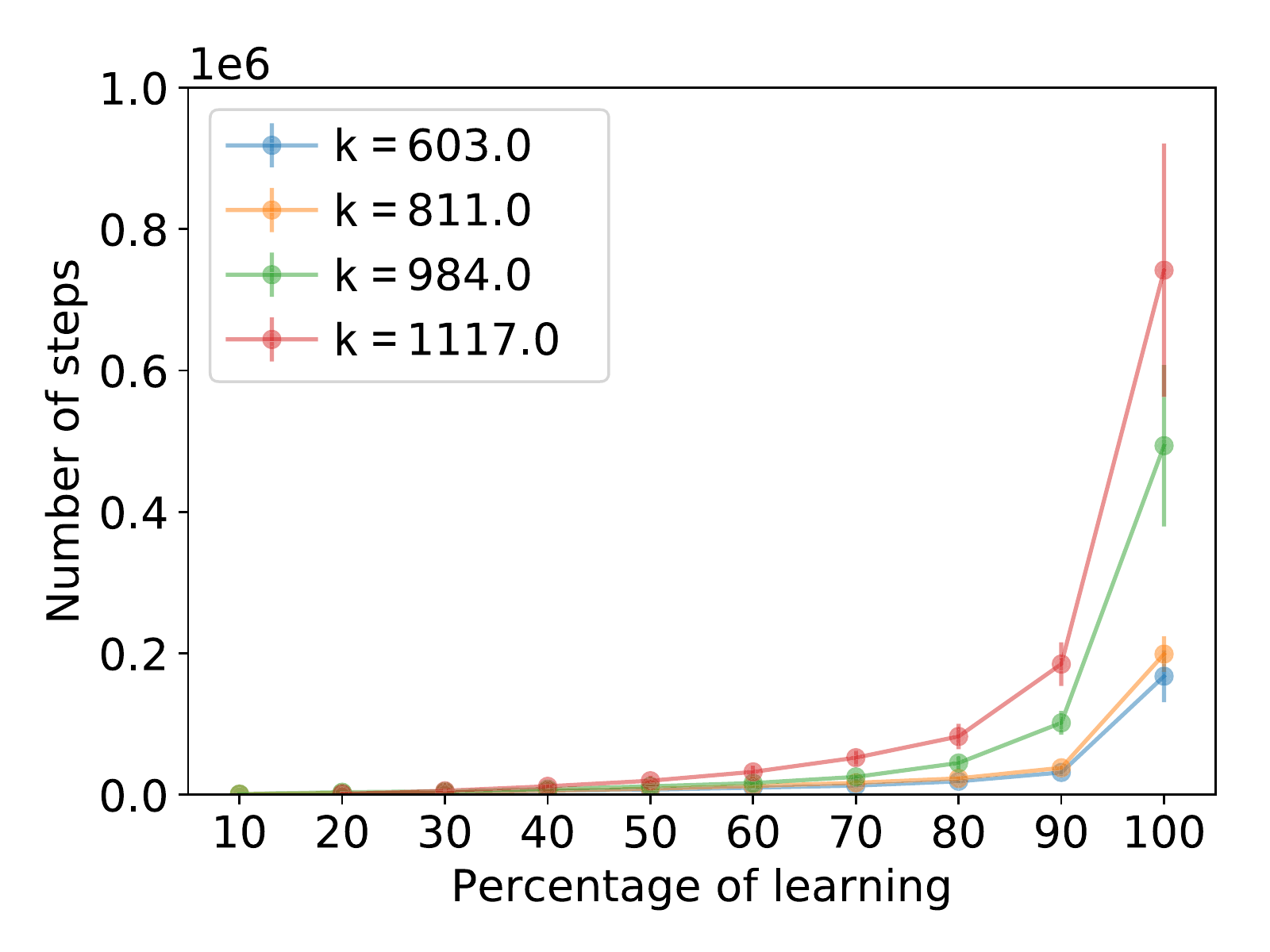}\\
  (a) $\langle k \rangle = 3$  & (b) $\langle k \rangle = 4$ & (c)  $\langle k \rangle = 5$
  \end{tabular}
   \caption{Curves of efficiency obtained by selecting the hubs as starting nodes in BA networks for the Look-ahead dynamics. The legends indicate the node degree of the starting node.}
  \label{hubs}
\end{figure*}

By selecting only the 4 nodes with greatest degrees for each network, we observe that the larger the degree, the less efficient is the discovery, as illustrated in Figure \ref{hubs}. This can be explained by the relatively high number of walks that start at the hub and terminate quickly. In the look-ahead dynamics, starting a walk at a hub tends to be highly inefficient since the agent checks every node around the hub.  The number of these nodes typically surpasses 10\% of the network. Moreover, since the known nodes are avoided, the chance of hitting a dead end is very high. This phenomenon can be seen in the video provided in the supplementary material.

\subsection{Structure Influence}

\begin{figure*}[!htb]
  \begin{tabular}{ccc}
  \includegraphics[width=.33\textwidth]{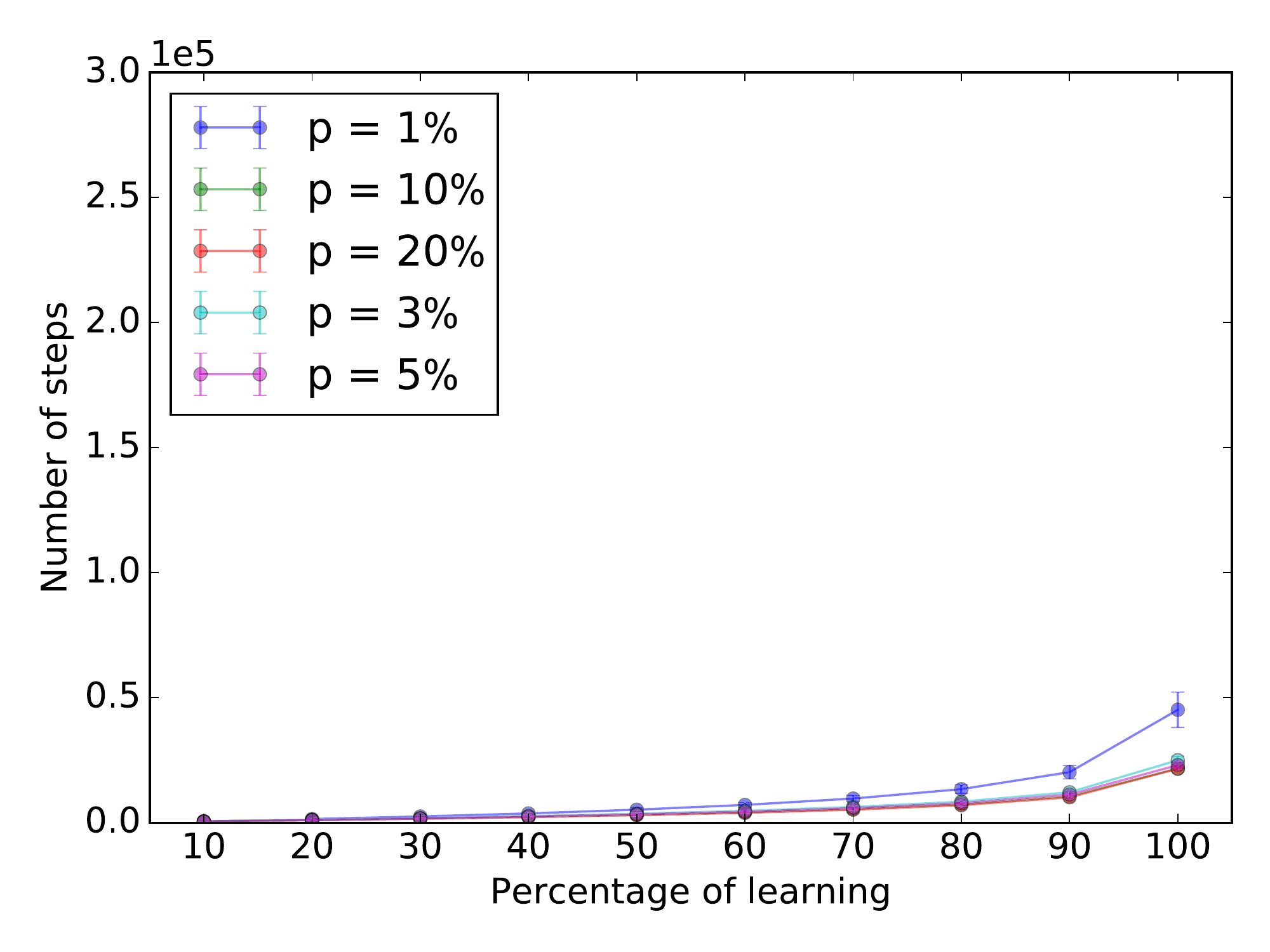} &   
  \includegraphics[width=.33\textwidth]{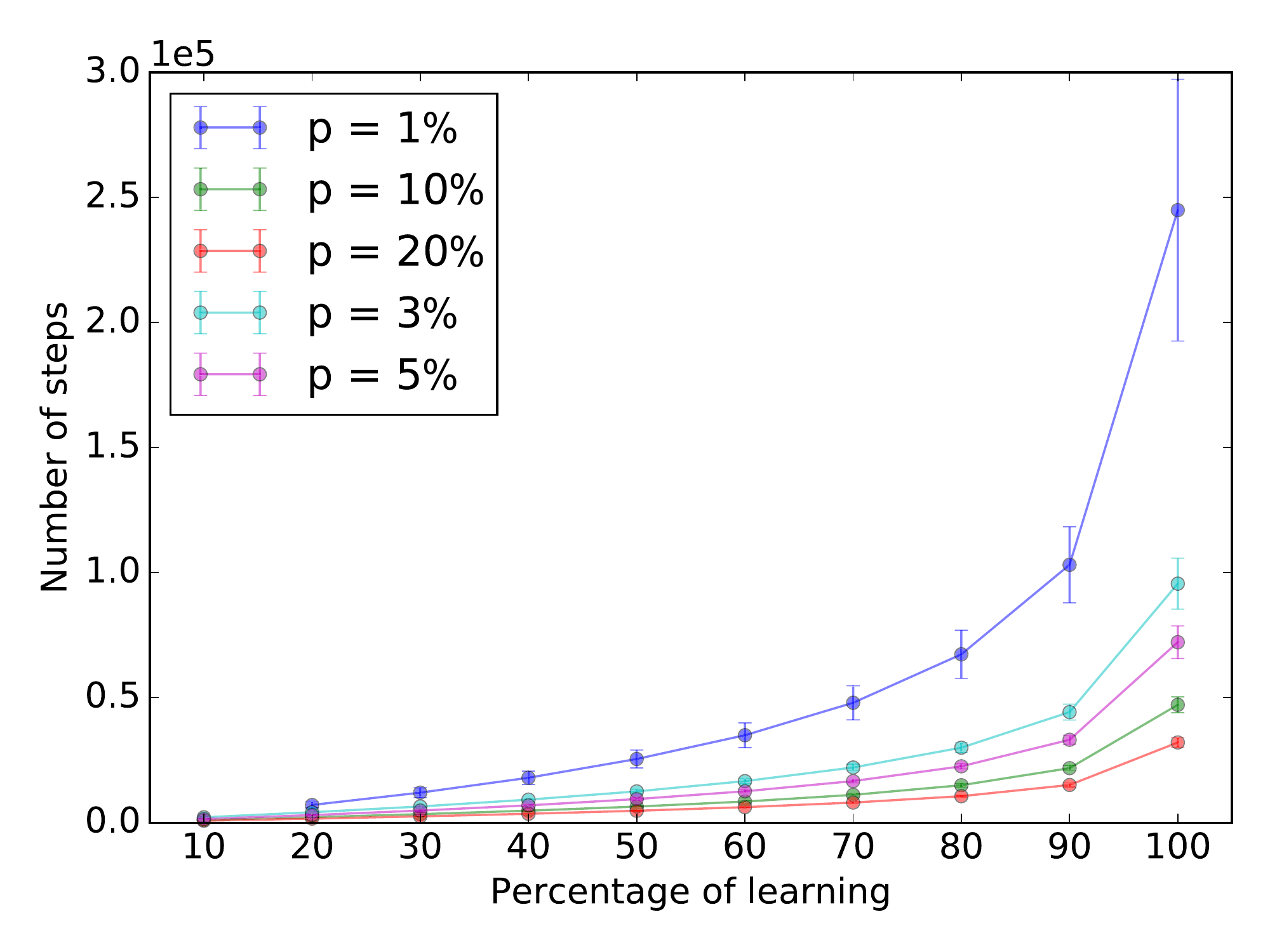} & 
  \includegraphics[width=.33\textwidth]{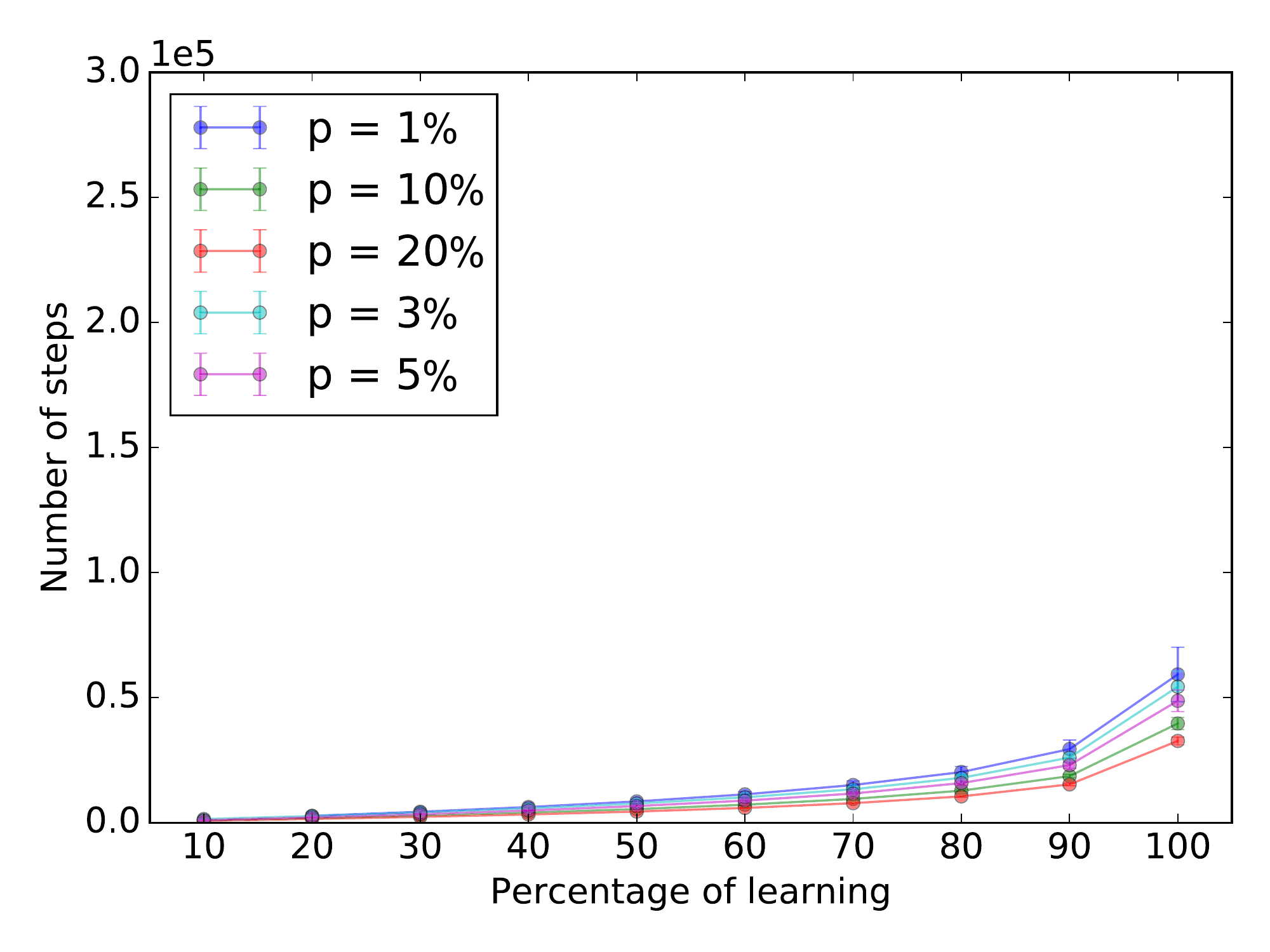}\\
  (a) WS Standard  & (b) WS Extended & (c)  WS Look-Ahead
  \end{tabular}
   \caption{Efficiency curves obtained for WS networks with varying rewiring probability $p$ and $\langle k \rangle = 4$.}
  \label{ws}
\end{figure*}

To analyze the effects of the structure on the efficiency of discovery process, we studied WS networks with varying probabilities $p$ of rewiring (see Figure~\ref{ws}). We can see that the extended self-avoiding walk has a much lower efficiency, which is manly because the neighborhood of a walk is highly overlapping due to the high clustering coefficient. Because of such a high local clustering observed in each step, the nodes around the starting node are counted several times. 
The standard walk does not pose such problem since it does not discover the nodes' neighbors and the look-ahead avoids known neighborhood thus generating very little overlap. 

\subsection{Real Network}
\begin{figure}[!hbtp]
  \centering
    \includegraphics[width=.4\textwidth]{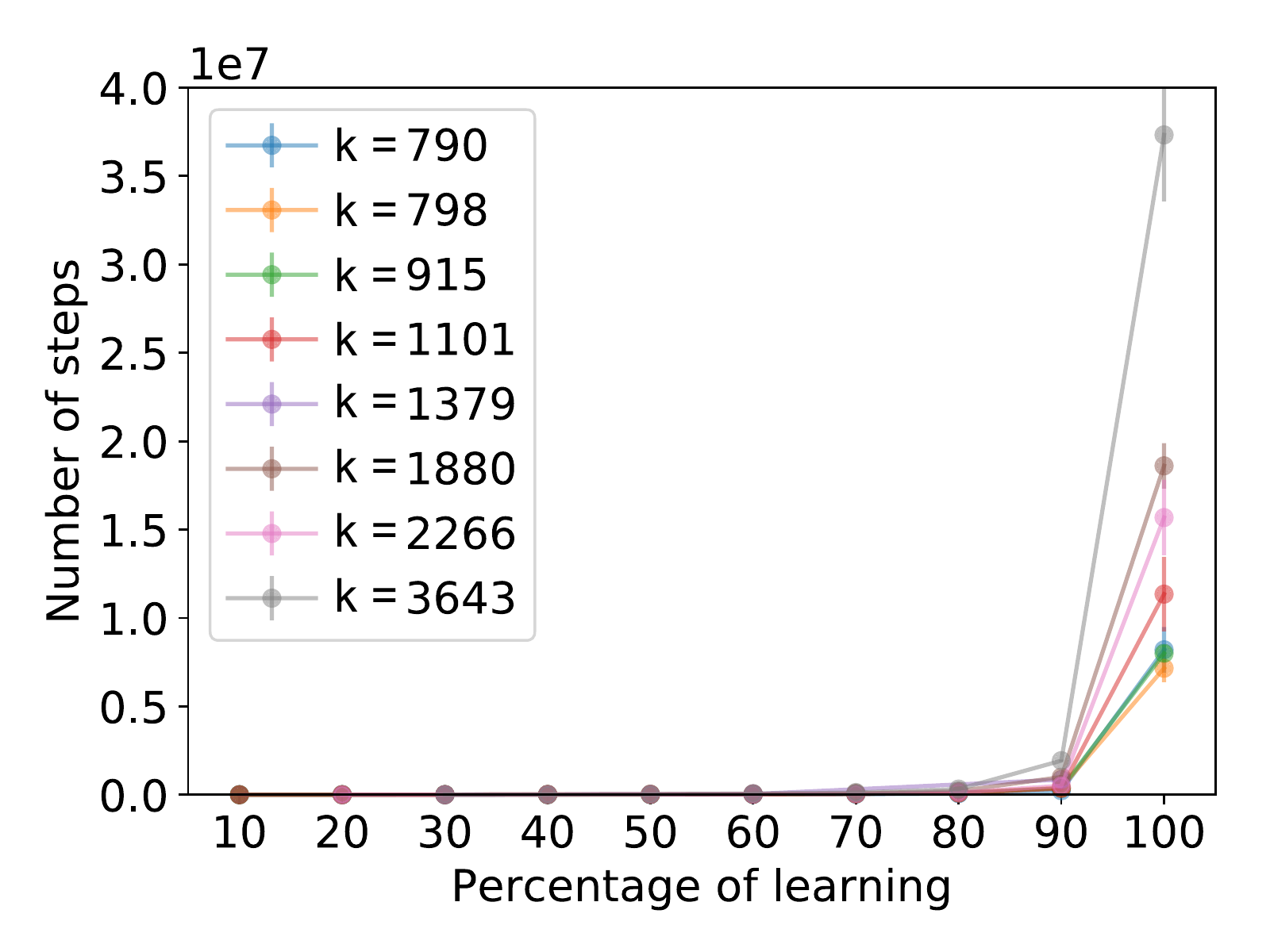}
   \caption{Learning curves obtained by simulations of the extended dynamics on the considered real-world network. The legends indicate the node degree of the starting node.}
  \label{realWalkTypes}
\end{figure}

The experiment using the considered real-world network (WOS) resulted in efficiency curves resembling those obtained for the BA and CM networks, as shown in Figure~\ref{realWalkTypes}. The selection of hubs as starting nodes was also found to play a major role in the dynamics. This corroborates the idea that influence of the starting node  on the dynamics is strongly related to its degree and that this behavior does not seems to depend on other, more complex, topological characteristics, which are present in the WOS network.

\section{Conclusions}
Better understanding how knowledge is acquired in several circumstances is of fundamental importance in science and technology, as it allows predictions to be made about the behavior of the studied systems, as well as contributes to make those processes faster and more accurate.  Being a particularly interesting and important type of knowledge acquisition, self-learning underlies several real-world systems, including human consciousness, risk assessment and prognostic, as well as the formation of social communities. Indeed, to a good extent, the rise of the WWW can be understood as a kind of self-knowledge.

While a good deal of works have investigated knowledge acquisition (see e.g.~\cite{arruda2017knowledge,KOPONEN2018405,doi:10.1093/comnet/cnu003,PhysRevLett.120.048301,LIN2010473,CAO2016277,ZHU2016192}), self-knowledge has received substantially less attention~\cite{WANG201783}. The present work approached this subject by using powerful concepts derived from network science. In particular, we investigated the self-knowledge acquisition as being a process derived from a random walk dynamics taking place in a information network. We considered three different types of dynamics and a diverse set of complex network models, as well as a real-world network representing the knowledge structure of a scientific area. By considering different combinations of network models and dynamics, three aspects were investigated in terms of acquisition efficiency: dependence on the dynamics, on the network type and on the local characteristics of the starting point.

The main contribution of this study is the result that the efficiency of self-knowledge acquisition, at least as approached from the point of view of the speed in which this dynamics unfolds and for the considered configurations and choices, hardly depends on either topology or learning strategy considered. This is a surprising result itself, as it hints that the several paths to self-knowledge are similar regarding their speed efficacy.

Similar results, though slightly more dependable on topology and dynamics, has been achieved in a previous investigation~\cite{arruda2017knowledge}, but the extension of this principle to the here reported self-knowledge is even more surprising because, unlike in that work, now the exploratory incursions are always performed with a fixed head-quarter, so that the topology surrounding that node would be expected to have a greater impact on the self-knowledge acquisition.

As future works, we intend to conduct a systematic study of the influence of topology, dynamics and starting node in a large variety of complex systems, including social~\cite{doi:10.1063/1.4960121}, information~ \cite{YIN2018935,ZHOU2018601,5871621}, semantical~
\cite{interplay,doi:10.1063/1.4954215} and technological networks~\cite{6544549}. We also intend to study whether some of the adopted models can reproduce the dynamics of acquisition and transmission of knowledge.




\acknowledgments

Thales S. Lima thanks Capes-Brazil for sponsorship. Henrique F. de Arruda acknowledges Capes-Brazil for sponsorship. Filipi N. Silva thanks FAPESP (grant no. 15/08003-4 and 17/09280-7) for sponsorship. Cesar H. Comin thanks FAPESP (grant no. 15/18942-8) for financial support. Diego R. Amancio acknowledges FAPESP (grant no. 16/19069-9 and 17/13464-6)
for financial support. Luciano da F. Costa thanks CNPq (grant no. 307333/2013-2) and NAP-PRP-USP for sponsorship. This work has been supported also by FAPESP grants 11/50761-2 and 15/22308-2.
\newpage
\bibliography{references}

\end{document}